\documentclass[twocolumn]{article}

\usepackage[letterpaper,margin=1in]{geometry}
\usepackage{newtxtext,newtxmath}
\usepackage{titlesec}
\usepackage{titling}
\usepackage{fancyhdr}
\usepackage[labelfont=bf]{caption}

\usepackage{amsmath, amsfonts}
 
\usepackage{amssymb}

\usepackage{graphicx}
\usepackage{setspace}
\usepackage{hyperref}
\usepackage{titlesec}
\usepackage{placeins}
\usepackage{multirow}
\usepackage[justification=raggedright, singlelinecheck=false]{caption}
\usepackage[
    backend=biber,
    style=authoryear,
    natbib=true,
    giveninits=true,
    terseinits=true,
    maxcitenames=2,
    mincitenames=1,
    maxbibnames=99,
    dashed=false,
    mergedate=false,
    isbn=false,
    doi=false,
    url=false,
    eprint=false,
    uniquename=false
]{biblatex}

\usepackage[normalem]{ulem} 
\DeclareFieldFormat{journaltitle}{\uline{#1}}
\DeclareFieldFormat{booktitle}{\uline{#1}}
\DeclareFieldFormat[article,inproceedings,thesis]{title}{\mkbibquote{#1\isdot}}

\DeclareNameAlias{default}{family-given}

\DeclareFieldFormat[article]{volume}{Vol.~#1}
\DeclareFieldFormat[article]{number}{No.~#1}
\DeclareFieldFormat{pages}{pp.~#1}

\renewbibmacro*{journal+issuetitle}{%
  \usebibmacro{journal}%
  \setunit*{\addcomma\space}%
  \iffieldundef{series}
    {}
    {\newunit
     \printfield{series}%
     \setunit{\addcomma\space}}%
  \printfield{volume}%
  \setunit{\addcomma\space}%
  \printfield{number}%
  \setunit{\addcomma\space}%
  \printdate 
  \setunit{\addcomma\space}%
}

\renewbibmacro*{issue+date}{}
\renewbibmacro*{note+pages}{%
  \printfield{pages}%
  \newunit}

\renewbibmacro{in:}{}

\usepackage{fancyhdr}
\fancypagestyle{symposium}{
    \fancyhf{} 
    \fancyhead[R]{\small 36th Symposium on Naval Hydrodynamics \\ Busan, Korea, June 14 –  19, 2026}
    
}

\titleformat{\section}{\bfseries\large}{\thesection}{1em}{}
\titleformat{\subsection}{\bfseries}{\thesubsection}{1em}{}

\title{\textbf{Reduced-Order Hydrodynamic Modelling of a Sphere Near a Wall Using Sparse Regression and Neural Networks}}

\author{\fontsize{14}{17}\selectfont
Z.\ Hoffman$^{1}$,
S.\ Vahaji$^{1}$,
A.\ Das$^{1}$,
M.\ Candon$^{1}$,
J.\ Nirman$^{2}$,
D.\ Sgarioto$^{3}$,
P.\ Marzocca$^{1}$ \\
\fontsize{14}{17}\selectfont
($^{1}$School of Engineering, RMIT University, Melbourne, Australia,\\
\fontsize{14}{17}\selectfont $^{2}$Navantia Australia, Melbourne, Australia,\\
\fontsize{14}{17}\selectfont $^{3}$Defence Science Technology Group, Melbourne, Australia)
}

\date{}

\begin{document}
\maketitle
\thispagestyle{symposium}
\begin{abstract}
This paper investigates a data-driven, reduced order method for real-time prediction of a floating sphere dropped near a vertical wall. The approach introduces a novel reduced-order model (ROM) formulation that sets \textit{a priori} boundaries to restrict the solution space for neural network learning. The method is demonstrated on a canonical one-degree-of-freedom heave-decay configuration. Experimental and analytical validation is used to inform the high-fidelity CFD simulation set up, which is then used to generate a parametric dataset consisting of 30 second motion trajectories spanning a range of wall distances (WD) and initial drop heights (DH). Sparse Identification of Non-linear Dynamics (SINDy) is applied to each CFD trajectory to identify the coefficients that describe a low-order non-linear ordinary differential equation (ODE) fluid oscillator. The SINDy-identified coefficients are then used as constraints in a neural network that learns a smooth mapping from WD and DH to the ODE coefficients, yielding a surrogate capable of predicting dynamics at arbitrary points in the input space without rerunning expensive CFD calculations. The surrogate reproduces CFD heave-decay responses with high accuracy while running in real-time. The approach provides a practical pathway toward real-time, physics-informed surrogate modelling for Launch and Recovery (LAR) operations.


\end{abstract}

\section{Introduction}
Launch and recovery of small uncrewed surface vessels (USVs) from larger parent vessels is a key enabler for future naval operations, but remains one of the most challenging phases of the mission. During launch and recovery, the USV is manoeuvred in close proximity to a much larger vessel, often in steep and irregular seas. A short-horizon, real-time forecasting tool for the USV during launch and recovery could support safer unmanned recovery across a wider range of sea states.

Conventional tools for predicting ship motions, such as strip theory and linear potential-flow methods, are well established and computationally efficient under the assumptions of inviscid, irrotational flow and small-amplitude motions \citep{Price1974}. In this linear regime, hydrodynamic forces are commonly decomposed into linear added mass, radiation damping, hydrostatic stiffness, and first-order wave-excitation components \citep{WAMIT, Han2024}. However, as wave steepness, motion amplitude, and geometric complexity increase, a linear description becomes progressively less accurate, as higher-order wave interactions, non-linear wave–radiation, and non-linear hydrostatic effects are no longer adequately captured \citep{Sulisz1993,IEAOES2024}. Similar limitations have been reported for ship--ship interactions in close proximity, where wave resonance in gaps and strong reflection effects introduce additional non-linearities \citep{Tan2017, Zhou2023b}. 

To resolve these effects with higher fidelity, recent studies have turned to Computational Fluid Dynamics (CFD), which can represent viscous effects, and non-linear free-surface dynamics at the expense of substantially increased computational cost \citep{Kramer2021}. However, because high-fidelity CFD remains too costly for real-time use, recent work in naval hydrodynamics has increasingly turned to surrogate modelling strategies capable of reproducing key non-linear behaviours at a fraction of the computational expense. 

Data-driven sequence models including LSTM \citep{Hao2022, Liong2024, Tang2021, Xu2021, Sun2022}, and hybrid CNN–RNN  architectures \citep{Cheng2021, Li2022, Zhang2019, Shi2025} have been applied to predict ship motions in waves showing some promising results for accurate real-time short-term motion prediction \citep{DAgostino2022, Silva2022}. Modal-decomposition approaches such as POD and DMD have also been used to construct low-dimensional representations that can be combined with neural networks for non-linear forecasting \citep{Francesco2021, Diez2024}. More recently, physics-informed (PINNs) \parencite{Schirmann2023, Guan2024} and operator-learning methods (Onets) \parencite{Zhao2025}, have been introduced to wave–body interaction problems to improve generalisation in data-sparse regimes. In parallel, sparse-regression techniques such as SINDy have shown that compact and interpretable non-linear ODEs can be recovered directly from time series dynamics data \citep{Brunton2016, Huang2023}.

Despite this progress, most existing surrogate models focus on isolated single-body dynamics. Comparatively little work has addressed the multi-scale, multi-body interactions characteristic of launch-and-recovery operations \parencite{Casey2019,Zhou2023b}. 

The methodology section of this paper is structed out as follows. First, the canonical heaving-sphere problem and the CFD-derived dataset used for model development is introduced. Then, a numerical validation is presented, followed by a physics-based dynamic models. The study is organised around a two-staged reduced-order modelling strategy. The SINDy implementation is described with discussions on the physical interpretation of the identified parametric coefficients. Finally, a neural-network regression maps the input parameters to these coefficients, producing a real-time surrogate that describes the sphere's dynamics over an arbitrary input.

\section{Background}
\subsection{Simplified LARS Configuration}
\begin{figure}[htbp]
    \centering
    \includegraphics[width=0.48\textwidth]{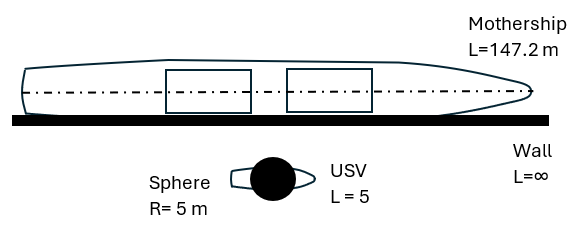}
    \caption{Schematic of the heaving sphere near a vertical wall, representing a simplified Launch and Recovery System (LARS) configuration.}
    \label{fig:sphere-geometry}
\end{figure}

\begin{figure}[htbp]
    \centering
    \includegraphics[width=0.48\textwidth]{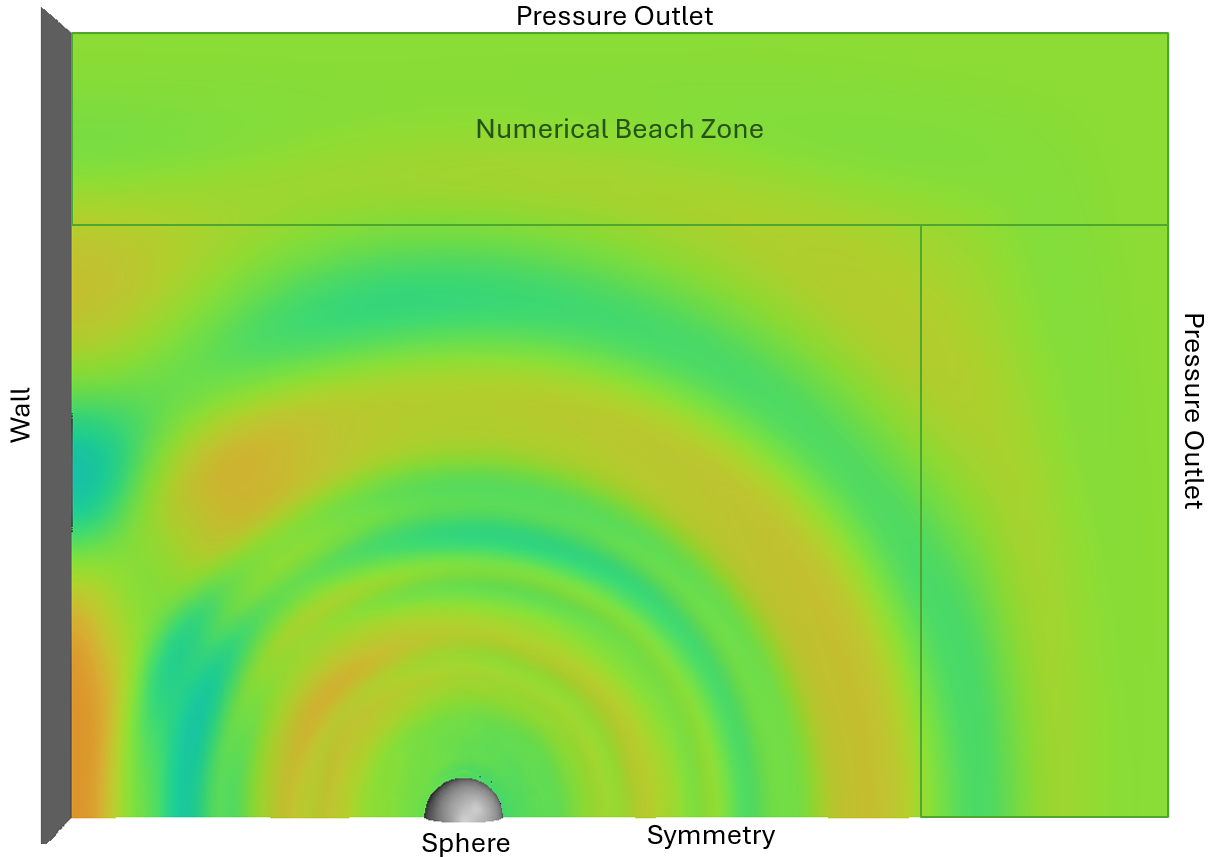}
    \caption{Freeze frame from CFD simulation.}
    \label{fig:CFD-freeze}
\end{figure}

In previous work, the authors developed and validated a CFD framework for a simplified LAR scenario in which a small USV is idealised as a one-degree-of-freedom heaving sphere operating near a large vertical wall representing the mothership shown in Figure \ref{fig:sphere-geometry} and \ref{fig:CFD-freeze}. This idealised problem is chosen because it retains the key hydrodynamic mechanisms relevant to launch and recovery while remaining well-supported by prior studies, including experimental measurements of sphere heave-decay motion \parencite{Kramer2021}, responses of a sphere under regular wave forcing \citep{IEAOES2024}, and analytically derived solutions \cite{Hulme1981}.

\subsection{CFD-Based Parametric Dataset}
\begin{figure}[h]
    \centering
    \includegraphics[width=0.49\textwidth]{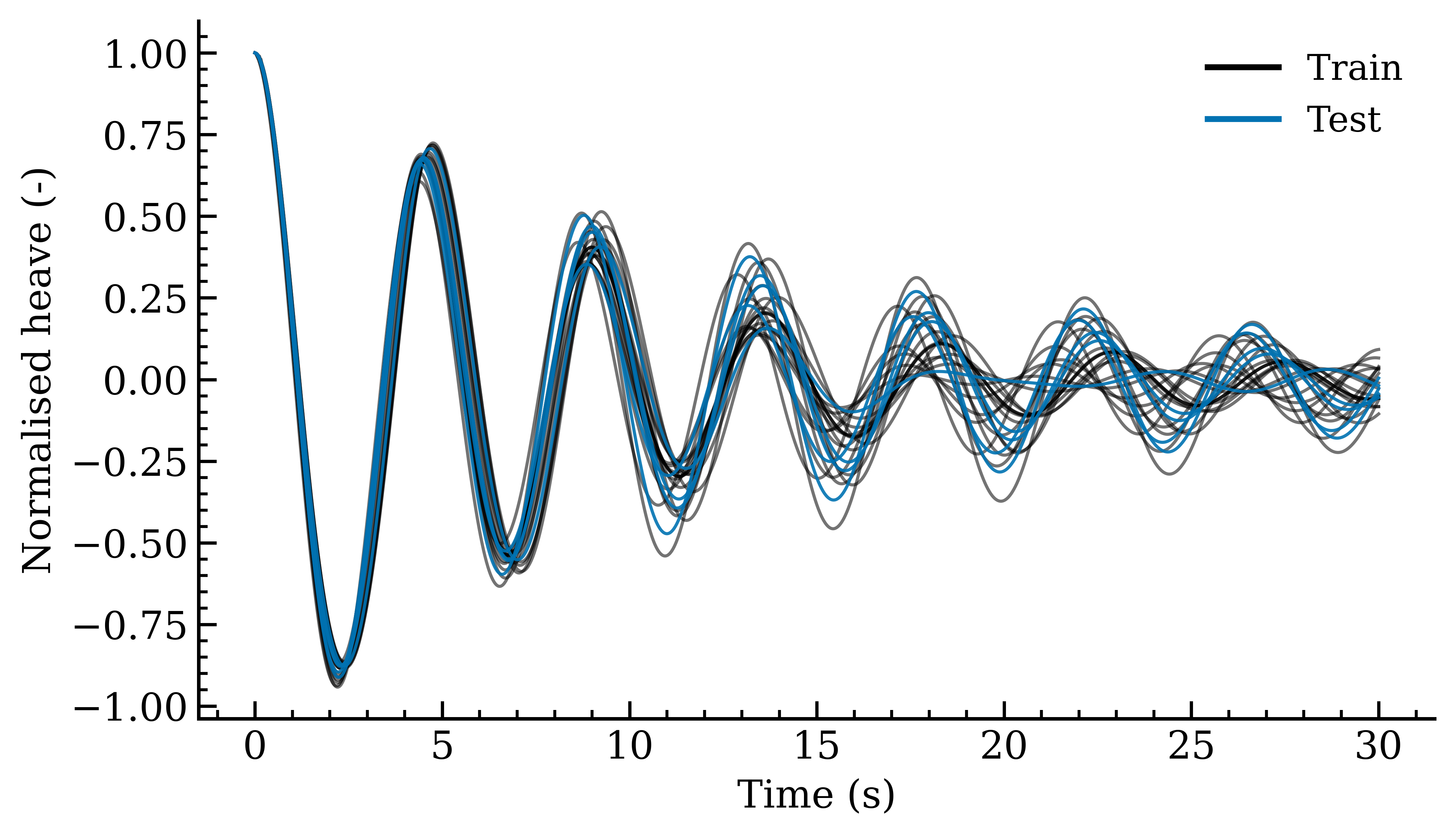}
    \caption{Raw dataset; train-test split}
    \label{fig:Train-test-split}
\end{figure}

\cite{Hoffman2025} have demonstrated that CFD can capture key non-linear hydrodynamic through comparing the simulation output with benchmark experimental data and analytical solutions. Based on the validated flow model, a parametric dataset of heave-decay responses across wall distances $WD \in (9,25)$m and drop heights $DH \in (0.5,5)$m was constructed. 
The drop heights considered are intentionally kept less than the sphere's radius to ignore physics relating to an entry-type problem. The heave response dataset, depicted in Figure \ref{fig:Train-test-split}, is normalised by dividing the initial heave position by the drop height ($\frac{x_{max}}{DH}$). A strong sensitivity on the input conditions ($WD,DH$) is evident. 

\subsection{Motivation}
While CFD is an appropriate tool for resolving the underlying physics of USV launch and recovery, it remains too expensive to employ directly for real-time usage. A single 30\,s heave decay simulation dropping a sphere into water can require $\mathcal{O}(10^2-10^3)$ CPU core-hours, and a comprehensive parametric study over WD and DH would demand thousands of such runs.

\section{Methodology}
\subsection{High Fidelity Hydrodynamic Modelling}
Ansys Fluent 2025R1 is used to numerically solve the Navier-Stokes for an isothermal and incompressible flow on a structured Cartesian grid. 

A dynamic overset mesh with 1-DoF heave motion is updated explicitly by integrating forces over the moving body’s surface. Flow variables are second order accurate in space and first order accurate in time $(CFL<0.3)$. A least square cell-based gradient and SIMPLEC pressure-velocity coupling are used. 

An explicit Volume of Fluid (VOF) approach in Equation \ref{eq:VOF} is adopted to capture the air-water interface using a Eulerian-Eulerian framework. In the continuity equation, the cells volume fraction $(0\le \alpha_{q} \le1)$ of the fluid phase $q$ are included:
\begin{equation}
    \frac{\delta}{\delta t}\alpha_q+\triangledown(\alpha_q \bar{v_q})=\sum_{p=1}^n(\dot{m}_{pq}-\dot{m}_{qp})
    \label{eq:VOF}
\end{equation}

Surface tension effects are neglected, justified by a Weber number well in excess of unity and under-relaxation factors for momentum, viscosity, and body forces are reduced $(n<1)$ to stabilise convergence. Two additional equations are solved for capturing the turbulence in the flow by employing the k-$\omega$ SST turbulence model. A turbulence damping source term was added to limit the over prediction of viscosity near the water-air interface, which arises from the VOF formulation whereby the turbulence models tend to over predict eddy viscosity from the high strain rate at the interface (\cite{ANSYS2025}).

Mesh and time-steps refinement studies were conducted to identify the coarsest discretisation that preserves solution accuracy.
\subsection{Linear Hydrodynamic Reference Model}
\label{subsec :linear_analytical }
\subsubsection{Decomposition of Hydrodynamic Forces}

For a heaving body, the total vertical hydrodynamic force can be decomposed into hydrostatic $F_{hs}$, radiation $F_{rad}$ and excitation $F_{exc}$ contributions, the wave excitation contribution can further be decomposed into the Froude-Krylov and diffraction force components.
\begin{equation}
F_{total} = F_{\mathrm{hs}}(x) + F_{\mathrm{rad}}(x,\dot{x}) + F_{\mathrm{exc}}(t),
\end{equation}
where $x(t)$ denotes the heave displacement. Assuming linearity, this leads to the familiar forced damped-spring mass system in Equation \ref{eq :linear_forces_general}
\begin{equation}
(M + A)\ddot{x}(t) + C\dot{x}(t) + Kx(t) = F_{\mathrm{exc}}(t),
\label{eq :linear_forces_general}
\end{equation}
with $M$ the structural mass, $A$ the added mass in heave, $C$ the radiation damping, and $K$ the hydrostatic stiffness i.e the buoyancy.

\subsubsection{Linear Properties of a Heaving Sphere}

For the spherical body used in this study, the structural mass and linear hydrostatic stiffness follow directly from the geometry,
\begin{equation}
M = \frac{2}{3}\rho_b \pi r^3,
\qquad
K_{linear} = \rho_{w} g \pi r^2
\label{eq :sphere_mass_stiffness }
\end{equation}
where $\rho_{w}$ is the water density, and $\rho_{b}$ is the spheres density chosen at $\rho_b=\frac{1}{2}\rho_w$, to match the experimental cases of \cite{IEAOES2024}. $g$ is gravitational acceleration, and $r$ is the sphere radius.

The linear added mass $A$ and damping $C$ coefficients for the heaving sphere are obtained analytically in the work of \cite{Hulme1981}.

\subsubsection{Heave Decay Response of the Isolated Sphere}

When the sphere is far from any wall and no incoming waves are present, the system reduces to a homogeneous spring--mass--damper oscillator driven only by linearly independent hydrostatics and radiation forces. The equation of motion becomes
\begin{equation}
    \bigl(M + A)\ddot{x}(t)+C\dot{x}(t)+Kx(t) = 0
    \label{eq :linear_decay_ode }
\end{equation}
The undamped natural frequency and (frequency-dependent) damping ratio are $\omega_0 = \sqrt{\frac{K}{M + A}}$ and $\delta(\omega) = \frac{C}{2\bigl(M + A\bigr)}$ respectively. 
 The resulting heave decay takes the standard damped-oscillator form $ \sqrt{\omega_0^2 - \delta(\omega_d)^2} $ where $C_1$ and $C_2$ are set by the initial displacement and velocity.

This gives the general solution to the ODE 
\begin{equation}
x(t)=(C_1cos(\omega _d t)+C_2 sin(\omega _d t)) e^{-\delta\omega _d t}
\end{equation}

\begin{figure}[htbp]
    \centering
    \includegraphics[width=0.43\textwidth]{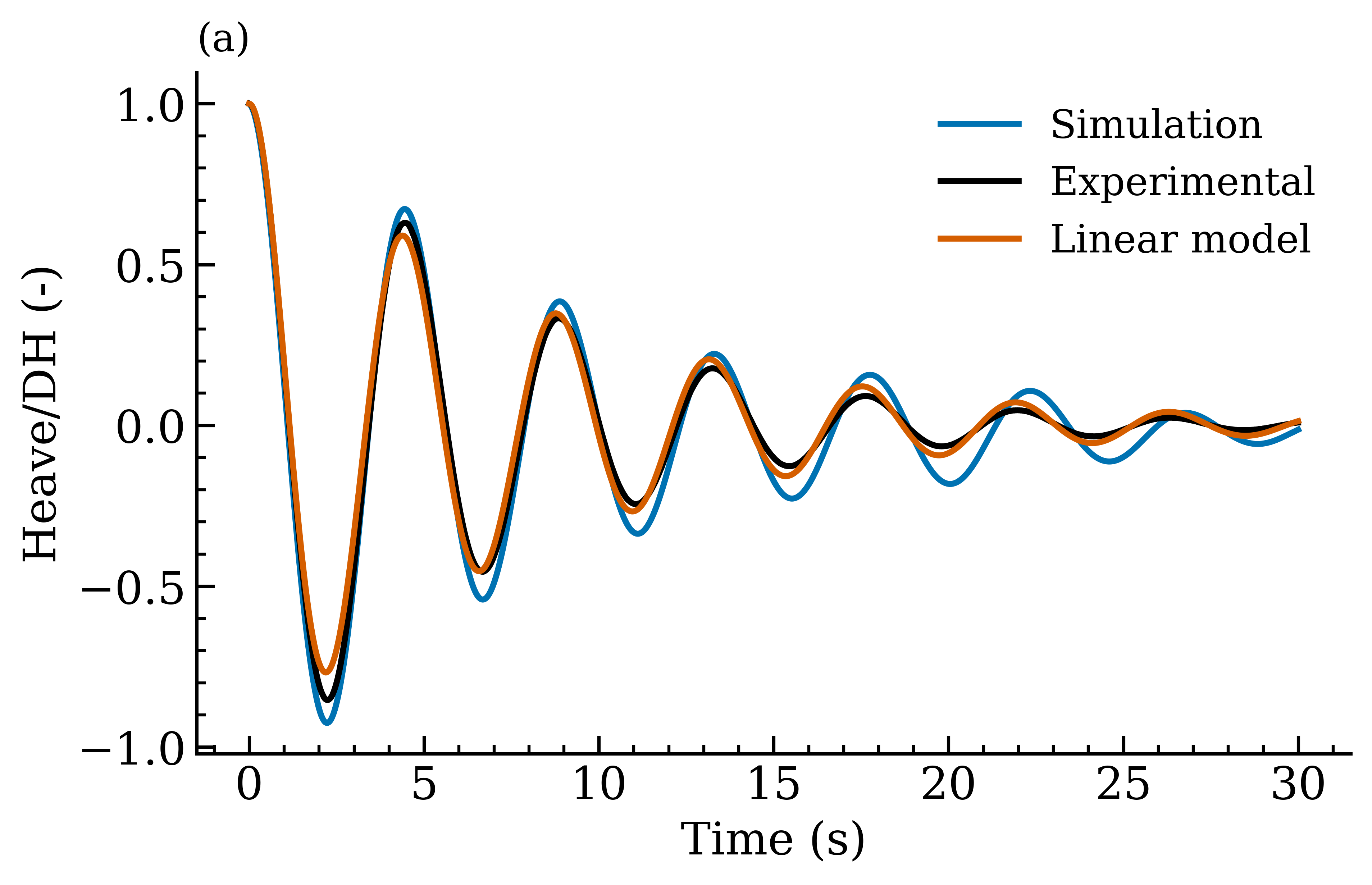}
    \includegraphics[width=0.43\textwidth]{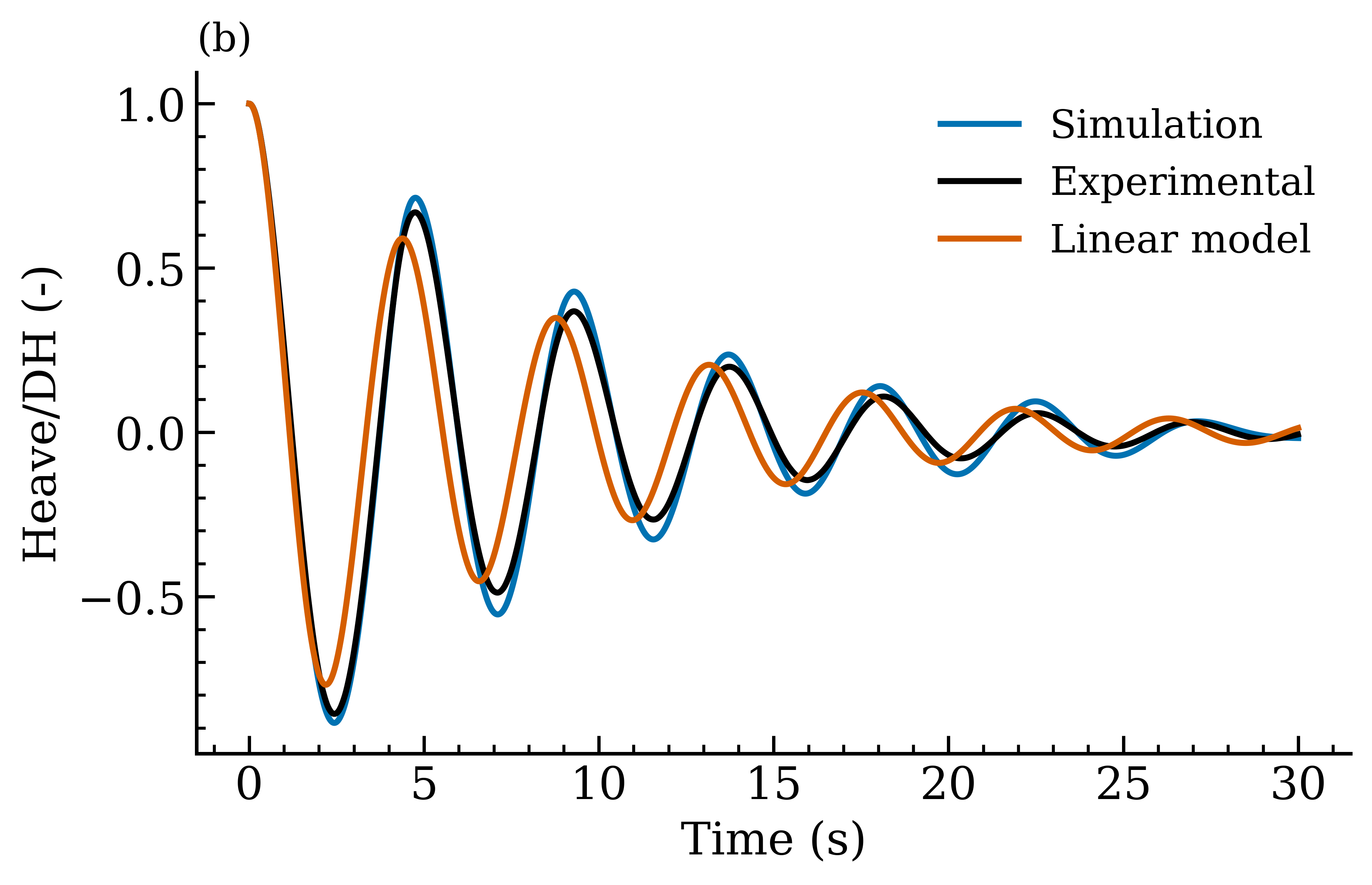}
    \caption{Validation of drop test from drop height of (a) $0.1$ and (b) $0.5$ of the spheres diameter. Comparison between our CFD simulation, experimental data \citep{Kramer2021} and linear model.}
    \label{fig:hv-decay-valid}
\end{figure}

For small drop heights, this linear model reproduces the CFD and experimental heave-decay envelopes with good accuracy (see Figure ~\ref{fig:hv-decay-valid}), confirming an adequate linear representation for small-amplitude motion. But, as the drop height increases, non-linear hydrostatic and radiation effects become significant, and the linear decay solution progressively deviates from the reference CFD, with the linear model under-predicting the peak amplitude and the decay frequency.  

\FloatBarrier 

\subsubsection{Sphere in Regular Waves: Linear Excitation Model}
When the sphere is subjected to regular incident waves, the motion can be captured in a linear ODE similar to that of a forced damped--spring--mass system. For a deep-water, first-order, linear Stokes wave theory, the free-surface elevation is given by:
\begin{equation}
\eta(t) = A_i \cos(\omega t),
\end{equation} 
and the corresponding wave-excitation force in heave is:
\begin{equation}
F_{exc}(t) = A_i \cos(\omega t) X,
\label{eq :linear_wave_force }
\end{equation}
where $A_i$ is the incident wave amplitude, and $\omega$ is the frequency of the incident waves. $X(\omega)$ is the frequency dependent excitation force coefficient per unit wave amplitude of a sphere, calculated using the commercial potential flow solver WAMIT in the work by \cite{Kramer2021}. 

The equation of motion becomes
\begin{equation}
\bigl(M + A)\ddot{x}
+B\dot{x}
+Cx
= F_{exc}(t)
\label{eq :linear_forced_ode }
\end{equation}
In a steady state, the linear solution has the form
\begin{equation}
x(t) = x_0 \cos(\omega t - \varphi),
\label{eq:forced-SS}
\end{equation}
where $x_0$ is the response amplitude and $\varphi$ is the phase lag. The associated response amplitude operator (RAO) in heave is
\begin{equation}
\mathrm{RAO}(\omega)
= \frac{x_0}{A_i}
\label{eq :linear_rao }
\end{equation}

Given analytically derived linear coefficients, the following linear model for the RAO is given in the paper by \cite{IEAOES2024}:
\begin{equation}
\mathrm{RAO}(\omega)
= \frac{X(\omega)}{\omega \sqrt{C^2+(\omega(M+A)-K/\omega)^2}}
\label{eq :rao_lin }
\end{equation}

\begin{figure}[t!]
    \centering
    \includegraphics[width=0.49\textwidth]{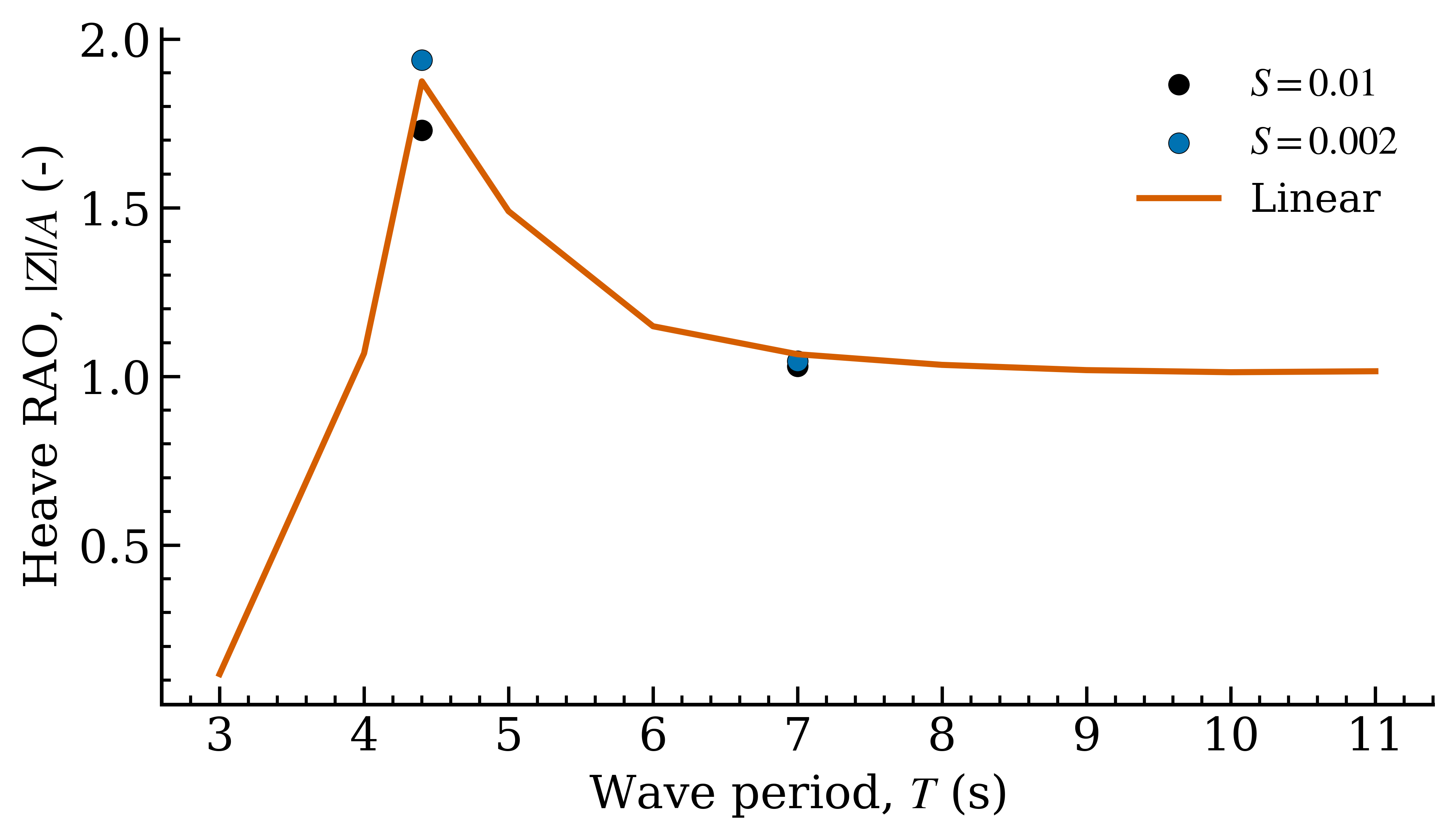}
    \caption{Validation of RAO for our CFD by comparing our results for the spheres heaving RAO with a linear model}
    \label{fig:Hv-RAO}
    
    \vspace{0.5cm} 

    
    \includegraphics[width=0.49\textwidth]{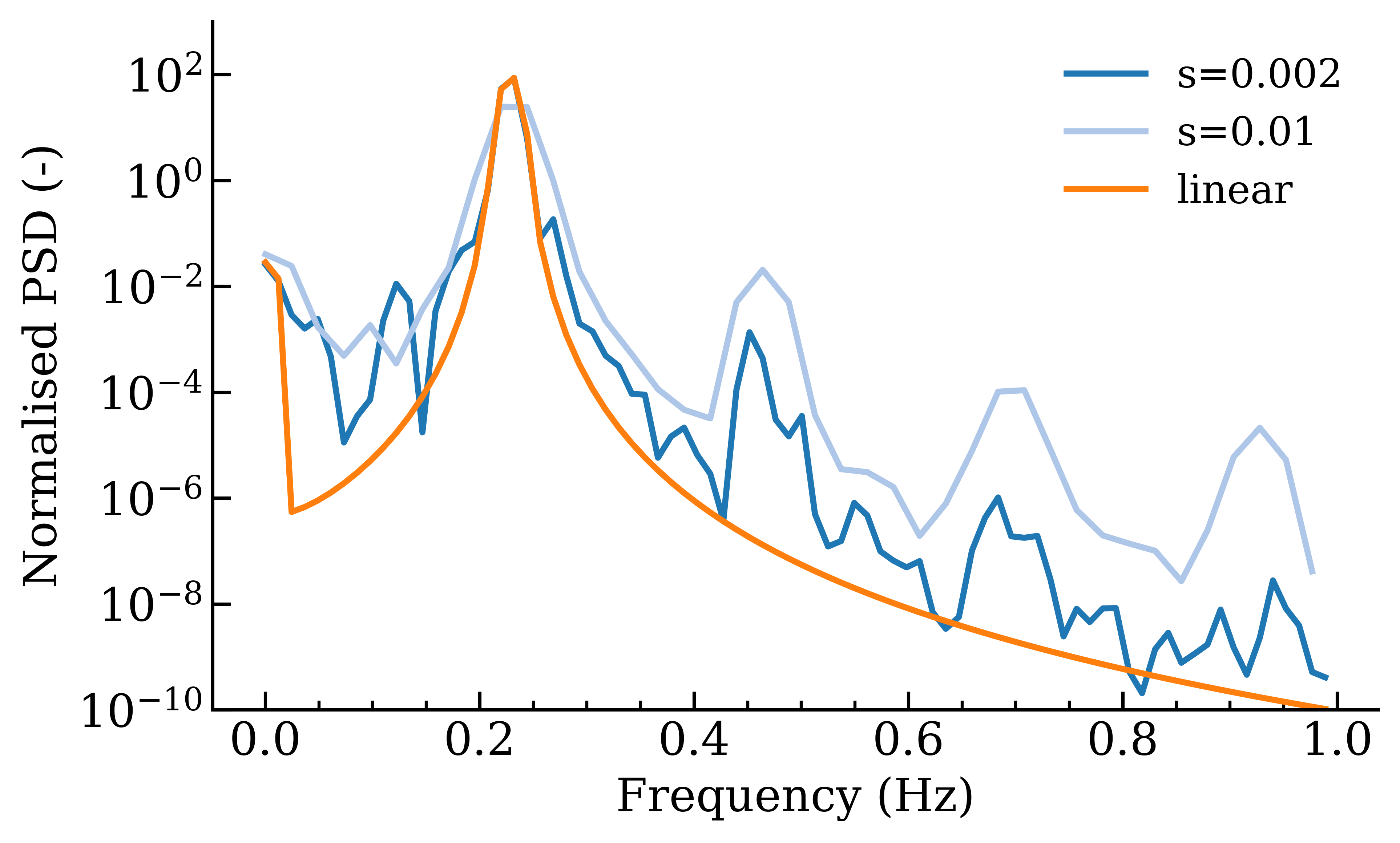}
    \caption{Normalised PSD of the sphere at resonance $T_{wave}=4.4$s for a steep wave $s=0.01$, a less steep wave $s=0.002$, and a linear response model}
    \label{fig:Hv-PSD}
\end{figure}

CFD and experimental RAOs are calculated in the validation study by taking the square root of the peak frequency identified in the power spectral density (PSD) of the heave response $S_{heave}(\omega)$ over that of the incident wave $S_{wave}(\omega)$. 
\begin{equation}
\mathrm{RAO}(\omega)
= \sqrt{\frac{S_{heave}(\omega)}{S_{wave(\omega)}}}
\label{eq :rao_data}
\end{equation}
Figure \ref{fig:Hv-RAO} shows a comparison between the linear model and the CFD derived RAOs. It is evident that the CFD results systematically depart from the linear model (Equation \ref{eq :rao_lin }) when the wave period is close to the spheres fundamental frequency and as wave steepness increase. This behaviour is consistent with the observations of \citet{Kramer2021}, who showed that higher-order hydrodynamic effects become increasingly significant with larger wave steepness $(s)$. Correspondingly, the PSD in Figure~\ref{fig:Hv-PSD} reveals a redistribution of energy from the fundamental frequency $\omega$ into higher harmonics (e.g.\ $2\omega$ and $3\omega$). As noted by \cite{Schoukens2019}, a linear time-invariant system cannot transfer energy between frequencies; thus, the appearance of these harmonics provides clear evidence of non-linear dynamical behaviour.

These discrepancies are indicative of non-linear diffraction force effects. In the CFD simulations, the instantaneous wave field around the sphere is strongly modified by the large body motion, the free-surface elevation, local waterline geometry and scattered wave patterns. This, we hypothesise, generates higher-order components in the scattered wave system and alters the effective excitation force. As a result, the true excitation cannot be represented by a single linear force coefficient $X$ relating the free surface $\eta$ to the excitation force $F_{exc}$.

\subsection{Non-linear Analytical Fluid Oscillator Model}

\subsubsection{Hydrodynamic Modelling for Launch and Recovery}
For launch-and-recovery scenarios several effects undermine linear assumptions. Large relative motions, steep waves, and geometric complexity introduce; higher-order wave interactions, non-linear changes in hydrostatic restoring force due to the instantaneous submerged volume, and non-linear excitation forces. Previous CFD and experimental studies of heaving bodies in waves have shown that these mechanisms lead to significant deviations from linear theory \citep{IEAOES2024,Sulisz1993}, whereby excitation and radiation forces are no longer linearly separable.

These observations motivate the present reduced-order modelling strategy, where, rather than abandoning the low-dimensional ODE representation, its coefficients are treated as non-linear functions the system-state $x$ and infer those functions directly from the CFD data.

A starting point is therefore to represent the dynamics using the following non-linear ODE of the form
\begin{equation}
(M + A)\,\ddot{x}+ C(x,\dot{x})\,\dot{x}+ K(x)\,x= F_{\mathrm{exc}}(t),
\label{eq:general_ode}
\end{equation}
where the effective added mass $A$, damping $C$, stiffness $K$, and excitation $F_{\mathrm{exc}}$ can be derived either analytically from first principles or from the raw data using linear regression based techniques. 

\subsubsection{Non-linear Hydrostatic Restoring Force}

\begin{figure}[htbp]
    \centering
    \includegraphics[width=0.5\linewidth]{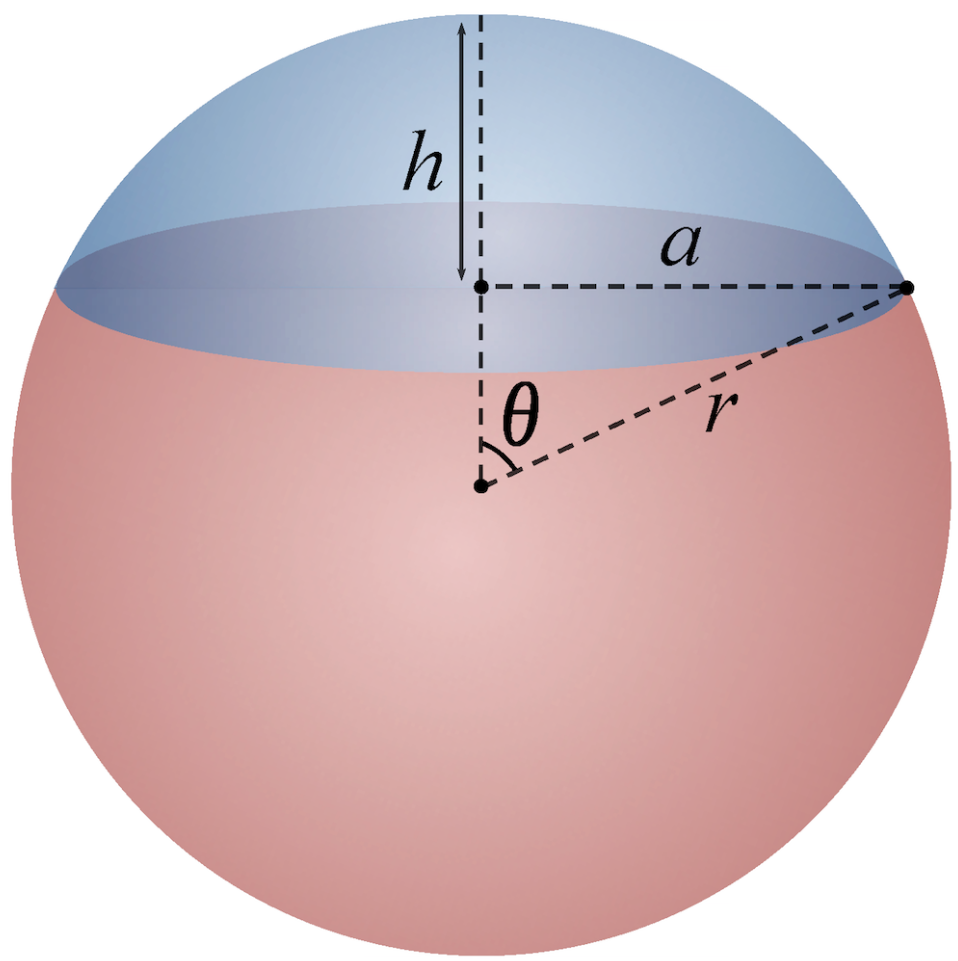} 
    \caption{Capped sphere geometry}
    \label{fig:Capped-sphere}
\end{figure}

For the capped sphere in Figure \ref{fig:Capped-sphere}, the exact hydrostatic restoring force can be written in terms of the volume as a function of height $V(h)$ such that;
\begin{equation}
V(h) = \frac{\pi}{3} h^2 (3r - h), 
\qquad h = DH\,x + r,
\end{equation}
where $R$ is the sphere radius, $DH$ is the initial drop height, and $x$ is the non-dimensional heave displacement (scaled by $DH$). 

\FloatBarrier

\begin{figure}[htbp]
    \centering
    \includegraphics[width=\linewidth]{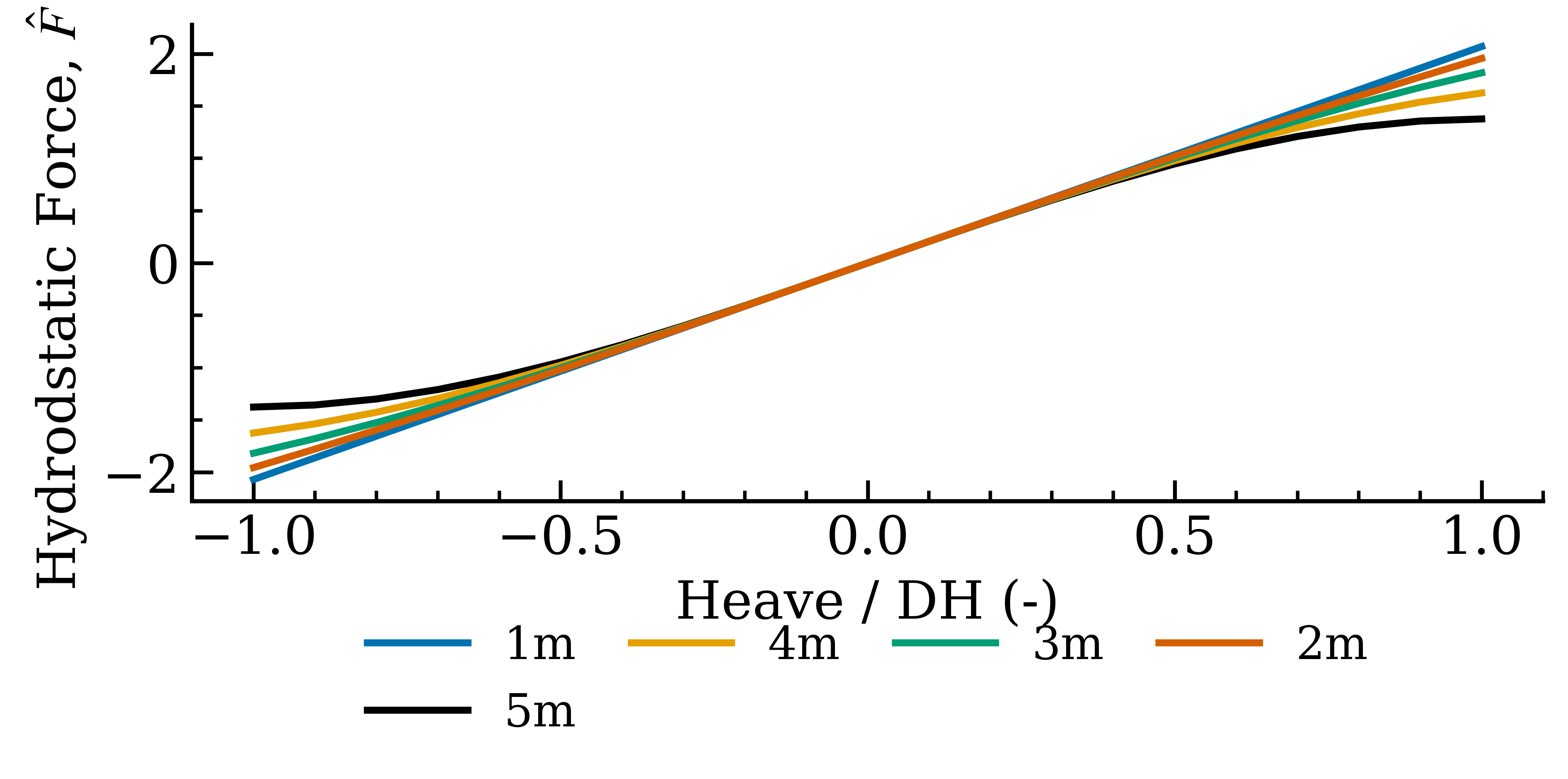} 
    \caption{Non-dimensional hydrostatic restoring force for the capped sphere, showing the cubic hardening behaviour with increasing displacement.}
    \label{fig:NL_hydrostatic_restoring}
\end{figure}

The vertical force is then normalised by drop height and mass in:
\begin{equation}
\hat{F}_{hs}=\frac{F_{hs}}{(M+A)\,DH}= \frac{V \rho g}{(M+A) DH},
\end{equation}
which, after substitution of $h=DH\,x+r$ and expansion, yields a hydrostatic relationship with a linear and cubic dependence on $x$. Writing the restoring term in the generic polynomial form
\begin{equation}
k(x,x^2,x^3)x = \Xi_1 x + \Xi_3 x^2 + \Xi_5 x^3,
\end{equation}
and matching coefficients gives the analytical approximation
\begin{equation}
k(x,x^2,x^3)\,x \approx 2.066\,x - 0.02755\,DH^2 x^3.
\end{equation}
The corresponding non-dimensional hydrostatic restoring curve is shown in Figure \ref{fig:NL_hydrostatic_restoring}, illustrates the departure from the linear approximation.

\FloatBarrier

\subsubsection{Numerical Time Integration of Non-linear ODEs}

To assess the predictive capability of our non-linear analytical model, the ODE is integrated in time and the resulting heave–decay response is compared against the CFD trajectories. The second–order equation
\begin{equation}
  \ddot{x}(t) = f\bigl(x,\dot{x},t\bigr)
  \label{eq:compact_ode}
\end{equation}
is recast as a first–order system by introducing the state vector
\begin{equation}
  \mathbf{s}(t) =
  \begin{bmatrix}
    x(t)\\[2pt]
    \dot{x}(t)
  \end{bmatrix},
  \qquad
  \dot{\mathbf{s}}(t)
  = \mathbf{f}\bigl(\mathbf{s}(t),t\bigr)
  =
  \begin{bmatrix}
    \dot{x}(t)\\[2pt]
    f\bigl(x(t),\dot{x}(t),t\bigr)
  \end{bmatrix}.
  \label{eq:RK4-SV}
\end{equation}
One explicit fourth–order Runge--Kutta (RK4) step from $t_n$ to
$t_{n+1}=t_n+\Delta t_n$ is then given by
\begin{align}
  \mathbf{k}_1 &= \mathbf{f}\bigl(\mathbf{s}_n, t_n\bigr),\\
  \mathbf{k}_2 &= \mathbf{f}\bigl(\mathbf{s}_n + \tfrac{\Delta t_n}{2}\mathbf{k}_1,
                                   t_n + \tfrac{\Delta t_n}{2}\bigr),\\
  \mathbf{k}_3 &= \mathbf{f}\bigl(\mathbf{s}_n + \tfrac{\Delta t_n}{2}\mathbf{k}_2,
                                   t_n + \tfrac{\Delta t_n}{2}\bigr),\\
  \mathbf{k}_4 &= \mathbf{f}\bigl(\mathbf{s}_n + \Delta t_n\,\mathbf{k}_3,
                                   t_n + \Delta t_n\bigr),\\
  \mathbf{s}_{n+1} &= \mathbf{s}_n
  + \frac{\Delta t_n}{6}\left(
    \mathbf{k}_1 + 2\mathbf{k}_2 + 2\mathbf{k}_3 + \mathbf{k}_4
  \right),
  \label{eq:RK4}
\end{align}
In all simulations a uniform time step of $\Delta t = 0.02\,\mathrm{s}$ is used, corresponding to $N = 1500$ steps, or $30\,\mathrm{s}$ in real time over the decay window.

\subsubsection{Comparison with Regression-based Non-linear Model}

Unlike the linear case, for which a closed-form singular, smooth analytical solution exists, the inclusion of non-linear hydrostatic and damping terms generally precludes a simple explicit solution to the ODE. Instead, the analytically derived non-linear restoring and damping terms are compared with those obtained by regression (SINDy-style least squares) from the CFD time series.

For a representative large drop height of $DH = 5\,\mathrm{m}$, the analytical model for the dominant terms in the acceleration reads
\begin{equation}
\ddot{x}_{\mathrm{analytical}}(DH=5\,\mathrm{m})
= -2.063\,x + 0.689\,x^3 - 0.24\,\dot{x},
\label{eq:dh-5m-an}
\end{equation}
whilst the corresponding regression-based model obtained from the raw CFD signal is
\begin{equation}
\ddot{x}_{\mathrm{regression}}(DH=5\,\mathrm{m})
= -2.057\,x + 0.686\,x^3 - 0.148\,\dot{x}.
\label{eq:dh-5m-sindy}
\end{equation}
The close agreement in the $x$ and $x^3$ coefficients confirms that the non-linear hydrostatic contribution is well captured analytically.

\subsubsection{Comparison Between Non-linear Analytical and Numerical}
\begin{figure}[h]
    \centering
    \includegraphics[width=0.9\linewidth]{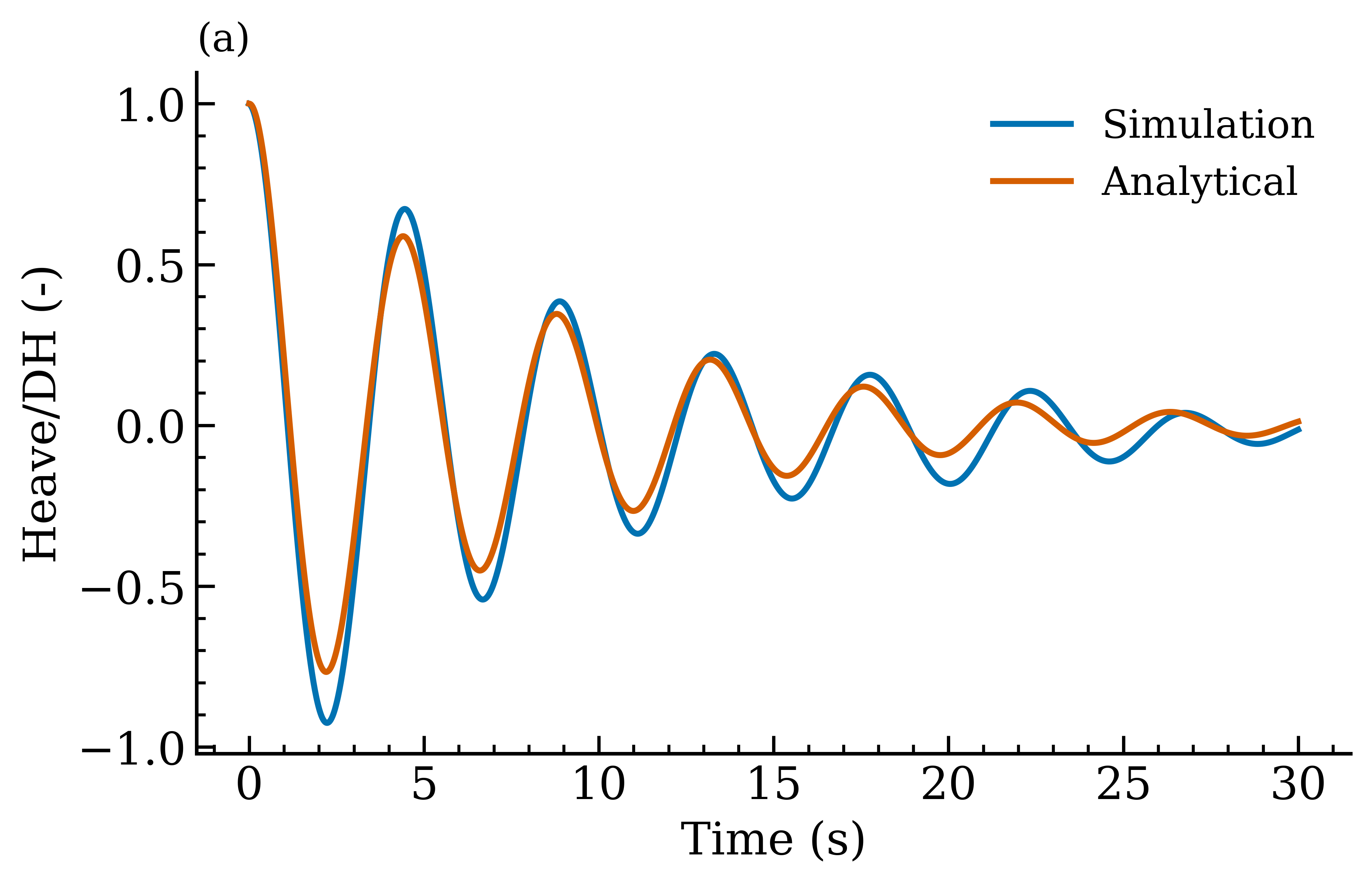}
    \includegraphics[width=0.9\linewidth]{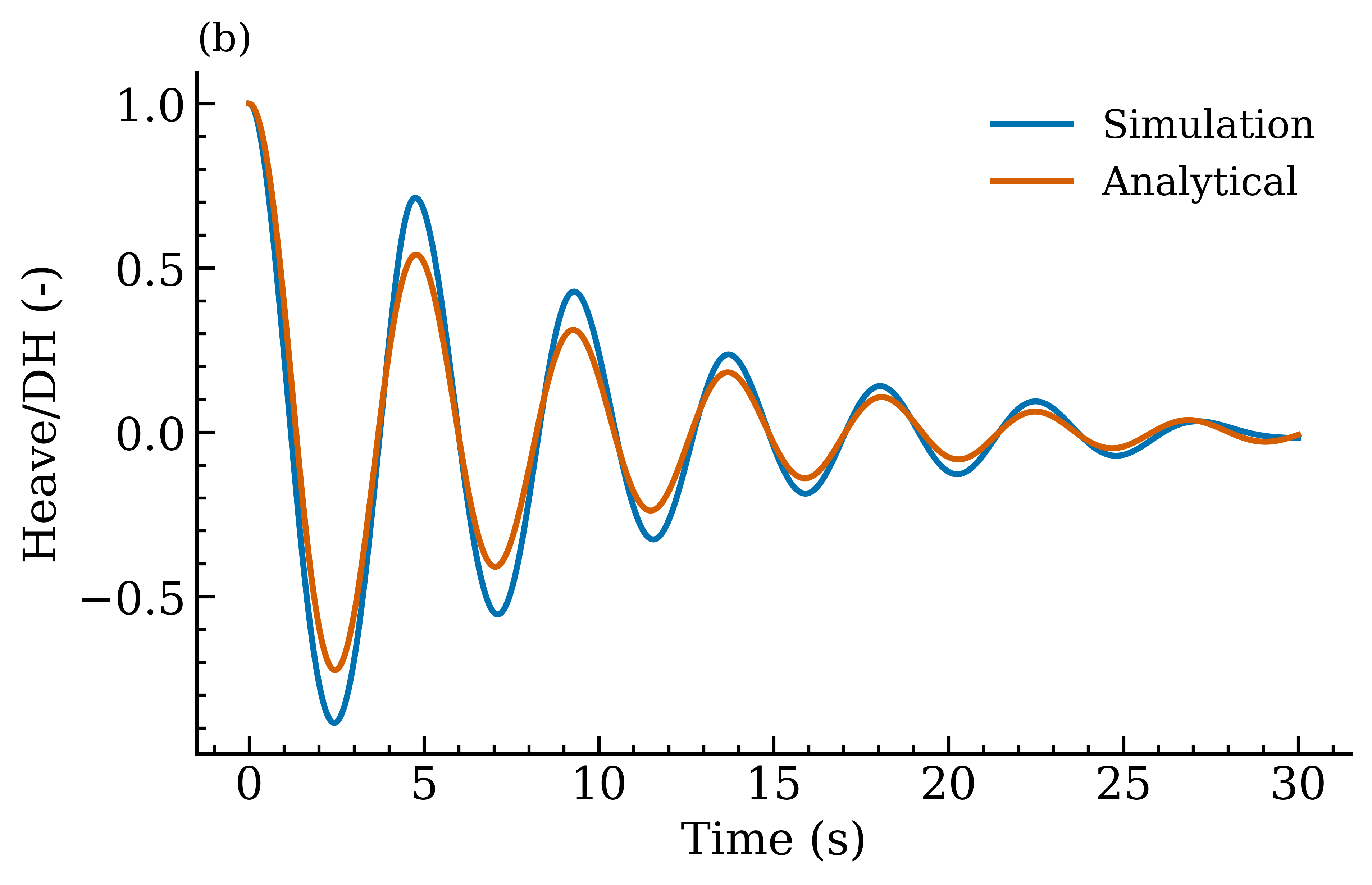}
    \caption{Heave decay for (a) $DH = 5\,\mathrm{m}$ and (b) $DH = 1\,\mathrm{m}$ comparison between raw CFD data and the non-linear analytical ODE model integrated with RK4.}
    \label{fig:Hv_comp_5m}
\end{figure}

Figure ~\ref{fig:Hv_comp_5m} compares the numerically integrated non-linear model with the raw CFD data for two representative drop heights. For small drops ($DH=1\,\mathrm{m}$), the response remains close to linear and the non-linear corrections are modest, whereas for larger drops ($DH=5\,\mathrm{m}$) (Equations \ref{eq:dh-5m-an} and \ref{eq:dh-5m-sindy}) the cubic hydrostatic term and amplitude-dependent damping play a dominant role in shaping the decay envelope and improving agreement with the CFD solution.

\subsection{Sparse Identification of Non-linear Dynamics (SINDy)}
Let's postulate that the normalised heave response satisfies a single-degree-of-freedom non-linear fluid oscillator of the form
\begin{equation}
\ddot{x}(t)
=
\sum_{k=1}^{6} \xi_k \,\Theta_k\!\left(x,\dot{x}\right)
+
\xi_7 \sin(\omega_0 t)
+
\xi_8 \cos(\omega_0 t),
\label{eq:fluid_oscillator}
\end{equation}
where $\Theta$ is the candidate library of terms, $\xi_1$ to $\xi_8$ are the coefficients of the parametric model, and $\omega_0$ is the natural frequency of the spheres heave. 

SINDy is the framework by which an ODE model can be recovered directly from time series data. $\ddot{x}$ can be represented as a sparse linear combination of candidate non-linear functions, allowing for the assembly of a candidate library matrix for the linear and non-linear terms forming our ODE;
\[
\Theta =
\begin{bmatrix}
x & \dot{x} & x^2 & \dot{x}^2 & x^3 & \dot{x}^3 & \sin(\omega t) & \cos(\omega t)
\end{bmatrix}
\]
and seek a coefficient vector $\Xi$ such that
\begin{equation}
    {\Xi}={\arg\min}\;
    \bigl\|\ddot{\mathbf{x}}(t) - {\Theta}{\Xi}\bigr\|_2^2
    +
    \alpha \bigl\|{\Xi}\bigr\|_2^2,
    \label{eq:ridge}
\end{equation}
where $\alpha$ is the two norm regularisation parameter. This system is solved by forming the equations in ridge regression such that
\begin{equation}
    \bigl({\Theta}^{T}{\Theta} + \lambda I\bigr){\Xi}
    = {\Theta}^{T}\ddot{x}
\end{equation}
a pseudoinverse $({\Theta}^{T}{\Theta} + \lambda I\bigr)^{-1} $ is taken based on singular value decomposition (SVD) \(\Theta = U\Sigma V^{T}\) to solve for the coefficients $\Xi$.

Sparsity is then enforced via Sequential Thresholded Least Squares (STLSQ) \citep{Huang2023}. At each iteration $k$:
\begin{enumerate}
    \item Identify ``small'' coefficients for $\Xi_k<\lambda$ where $\lambda$ is a sparsity threshold.
    \item Set ${\Xi}_k = 0$ and refit Equation \ref{eq:ridge} using only the remaining ``active'' columns of ${\Theta}$.
    \item This process is repeated until the active set stabilises.
\end{enumerate}
The coefficients are subsequently mapped back to the original, de-normalised feature space, yielding a sparse vector $\Xi$ associated with the functions in Equation \ref{eq:fluid_oscillator}.

Here, the SINDy-identified terms can be grouped into physically meaningful contributions. For example, the polynomial terms represent non-linear hydrostatic and radiation forces, while the sinusoidal entries are a crude representation of the wave excitation force arising from the interaction between the sphere and the rebounded waves. 

\begin{figure}[htbp]
    \centering
    \includegraphics[width=\linewidth]{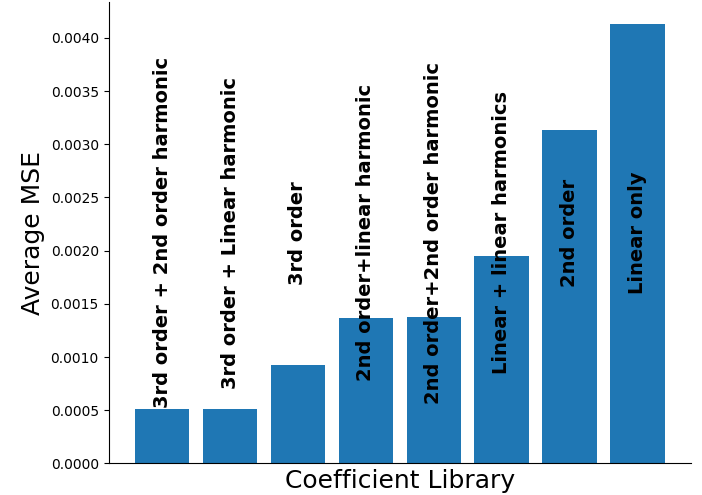} 
    \caption{The required non-linear terms include in the library to best capture the dynamics (MSE).}
    \label{fig:Coefficient-library}
\end{figure}

To determine an appropriate candidate library, SINDy is applied separately to each $(WD,DH)$ case by solving a regularised least-squares problem and evaluated by taking the average MSE over the entire dataset. Figure \ref{fig:Coefficient-library} shows the candidate library evaluated in a \texttt{for}-loop over increasing polynomial and harmonic orders, which allows us to identify the best set of terms required to obtain a satisfactory representation of the dynamics

After evaluating the accuracy of several candidate libraries, the optimal model was identified as the third-order polynomial plus first-order harmonic library in Equation \ref{eq:fluid_oscillator}. This basis is both compact and sufficiently accurate, achieving an error of $MSE=5.1\times 10^{-4}$. Including higher-order harmonics did not yield any improvement in accuracy; a better model would require a more realistic the excitation-force model.
\FloatBarrier

\subsubsection{Excitation Force Model}
\begin{figure}[htbp]
    \centering
    \includegraphics[width=\linewidth]{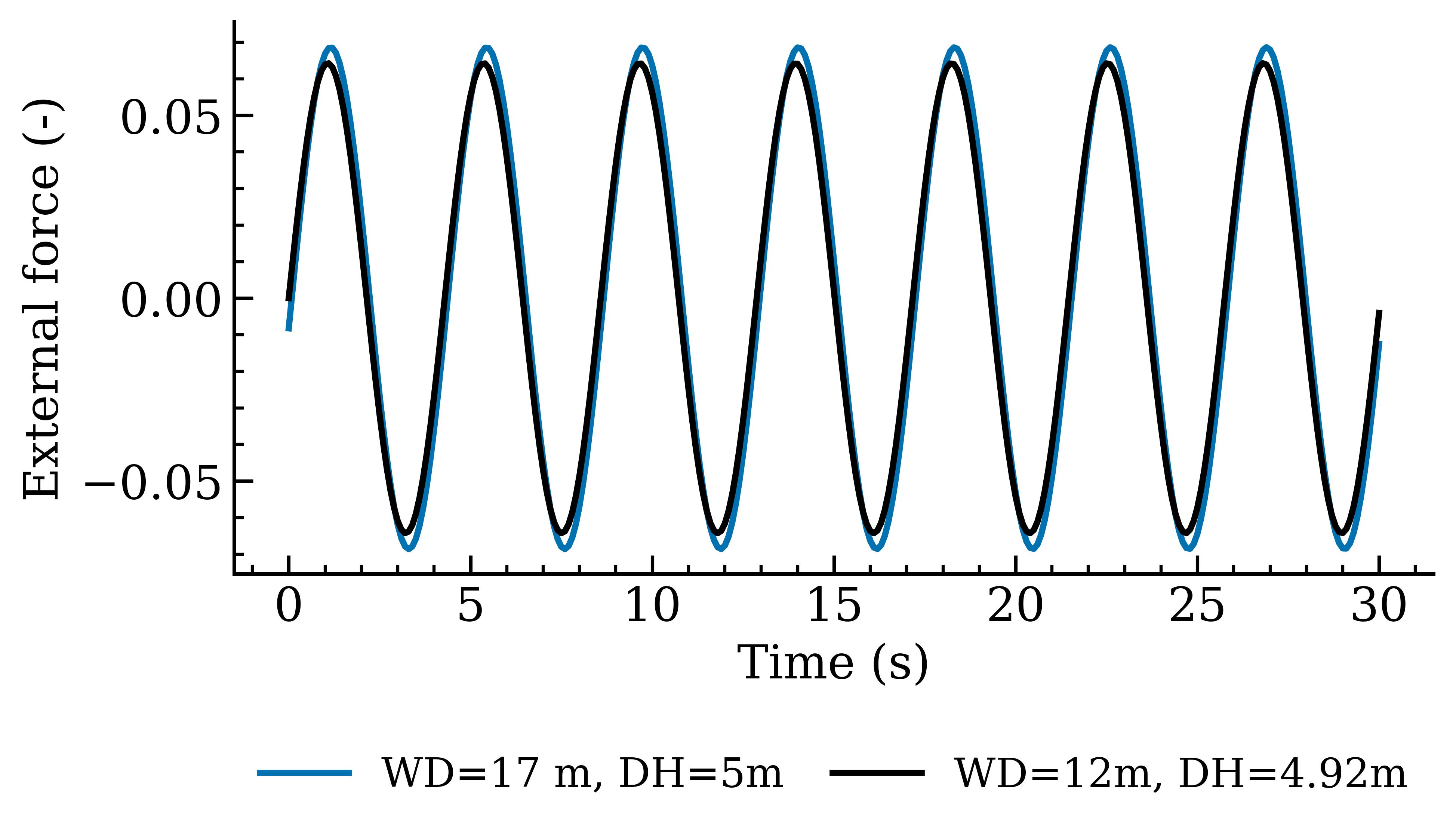} 
    \caption{Simplified excitation force model}
    \label{fig:Diff-simple}
\end{figure}

To represent the wave excitation forces from the waves rebounding from the wall, the SINDy feature library is augmented with single–frequency harmonic terms in Figure \ref{fig:Diff-simple}, such that:
\[
\Theta \;\supset\;
\begin{bmatrix}
\sin(\omega t) & \cos(\omega t)
\end{bmatrix},
\]
which act as a simple, low–order surrogate for the excitation force.  

From the identified coefficients of $\sin(\omega t)$ and $\cos(\omega t)$, SINDy provides an estimate of the effective excitation amplitude and phase. This estimate is useful in capturing the general magnitude and phase of the excitation force, but as it remains a crude approximation, it is incapable of fully capturing the true physics.

To investigate whether the wall-induced excitation could be isolated, an additional CFD simulation was performed for the same drop condition with no nearby wall. This wall-free case was used to estimate the homogeneous force contribution, \(F_{\mathrm{homogeneous}}\). It was then assumed, that the force measured in the case with the wall-present could be decomposed into a homogeneous contribution and an additional reflected wave excitation contribution,
\[
F_{\mathrm{total}}(t)
=
F_{\mathrm{homogeneous}}(t)
+
F_{\mathrm{exc}}(t).
\]

\begin{figure}[htbp]
    \centering
    \includegraphics[width=\linewidth]{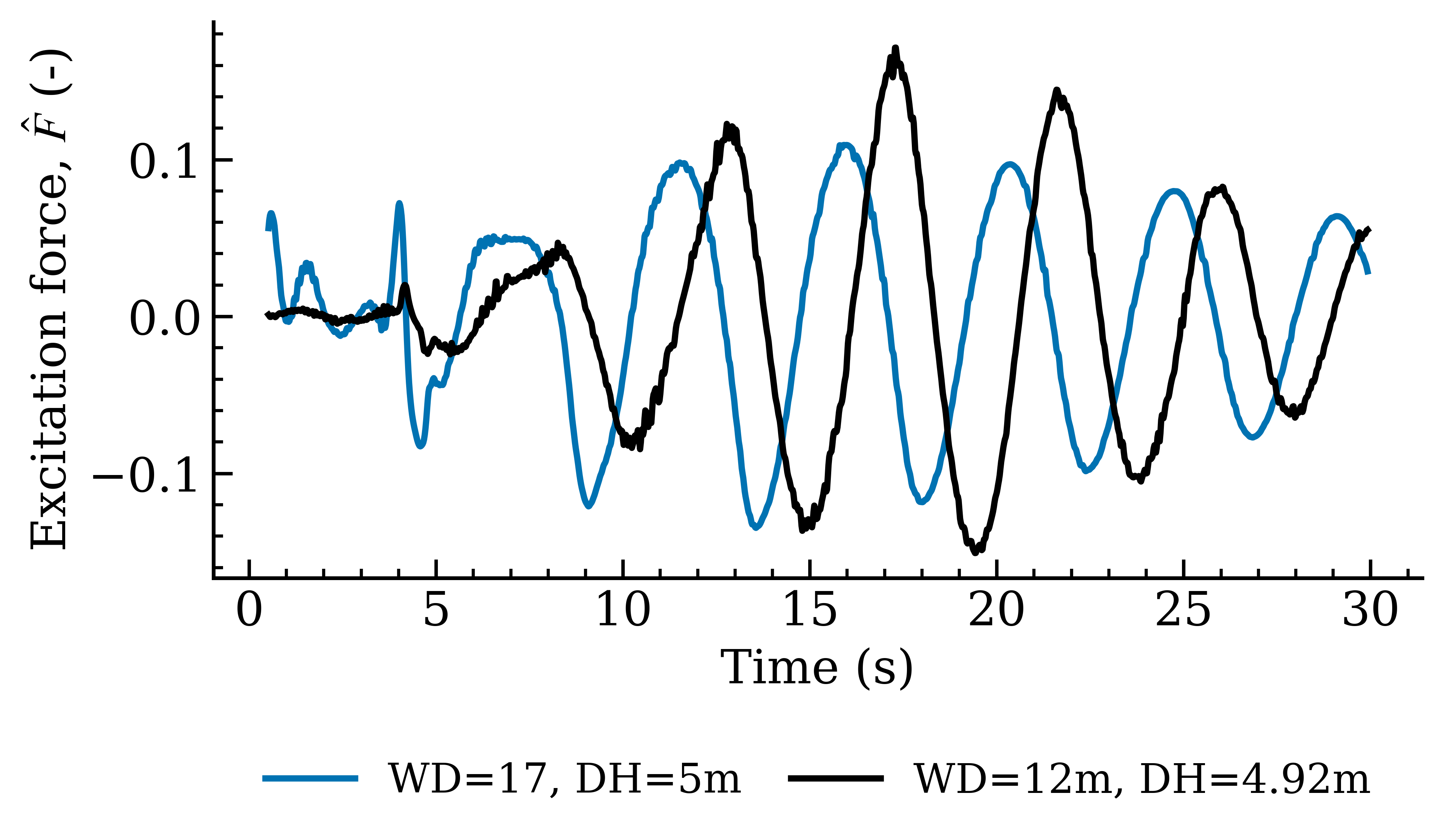} 
    \caption{Linear excitation forces assumption}
    \label{fig:Diff-lin}
\end{figure}

Under this assumption, the normalised excitation force ($\hat{F}=\frac{F}{(M+A)DH}$) was estimated by subtracting the wall-free force from the force when the wall is present:
\[
\hat{F}_{\mathrm{exc}}(t)
=
\hat{F}_{\mathrm{total}}(t)
-
\hat{F}_{\mathrm{homogeneous}}(t),
\]
leading to the reconstructed excitation shown in Figure~\ref{fig:Diff-lin}. Under this assumption, the homogeneous dynamics for a representative case (e.g.\ $DH = 5\,\mathrm{m}$) can be written as
\begin{equation}
\ddot{x}+ 2.057\,x - 0.686\,x^3 + 0.148\,\dot{x} = 0,
\end{equation}
and one might hope to write a non-linear ODE representing the dynamics of the sphere with the returning waves from the wall as 
\begin{equation}
\ddot{x} + 2.057\,x - 0.686\,x^3 + 0.148\,\dot{x} = F_{\mathrm{exc}}(t).
\end{equation}
such that
\begin{equation}
F_{\mathrm{exc}}(t)=X\eta(t)
\end{equation}

\begin{figure}[htbp]
    \centering
    \includegraphics[width=\linewidth]{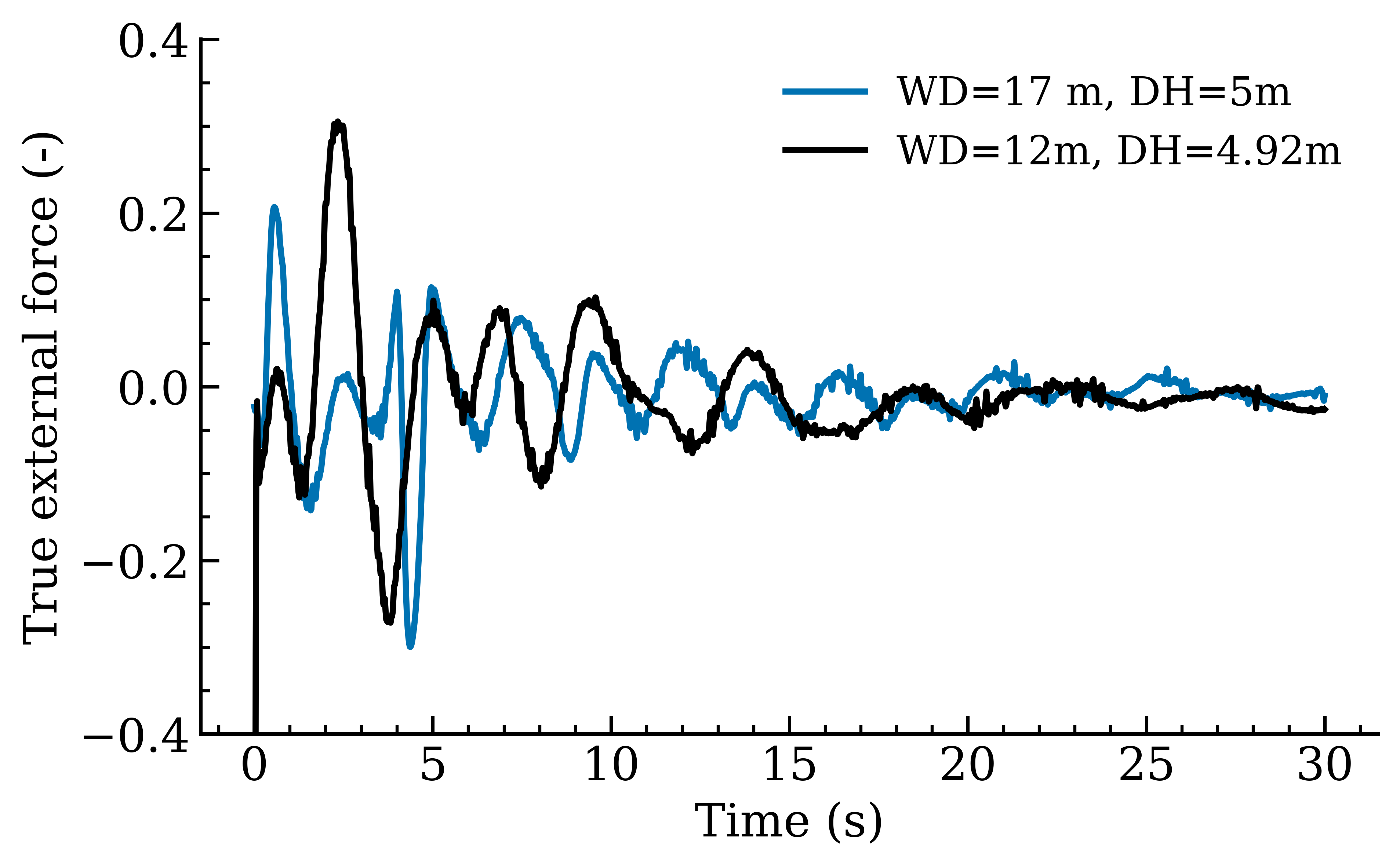} 
    \caption{True excitation forces}
    \label{fig:Diff-True}
\end{figure}

In practice, this linear superposition fails such that: 
\begin{equation}
(M+A)\ddot{x}+C(x,\dot{x})\dot{x}+Kx \neq X\eta(t)
\end{equation}
The excitation force $F_{\mathrm{exc}}(t)$ inferred in this way does not reproduce the CFD heave time history when re–integrated, and the true excitation force $F_{exc}$ required to recover the true trajectory is shown in Figure~\ref{fig:Diff-True} differs markedly from Figure \ref{fig:Diff-lin}. This indicates that excitation force is non-linearly coupled to the body motion and radiation field, so that the true excitation cannot be represented by a single linear term that is independent of the state. In future work, explicitly considering the Froude-Krylov and radiation forces separately may be considered to potentially provide an improved model of the excitation forces.

Despite this lack of linearity, there is potentially some information on the excitation force that can be recovered in the simplified sin/cosine coefficients determined using our SINDy regression.

\begin{figure}[t]
    \centering
    \includegraphics[width=\linewidth]{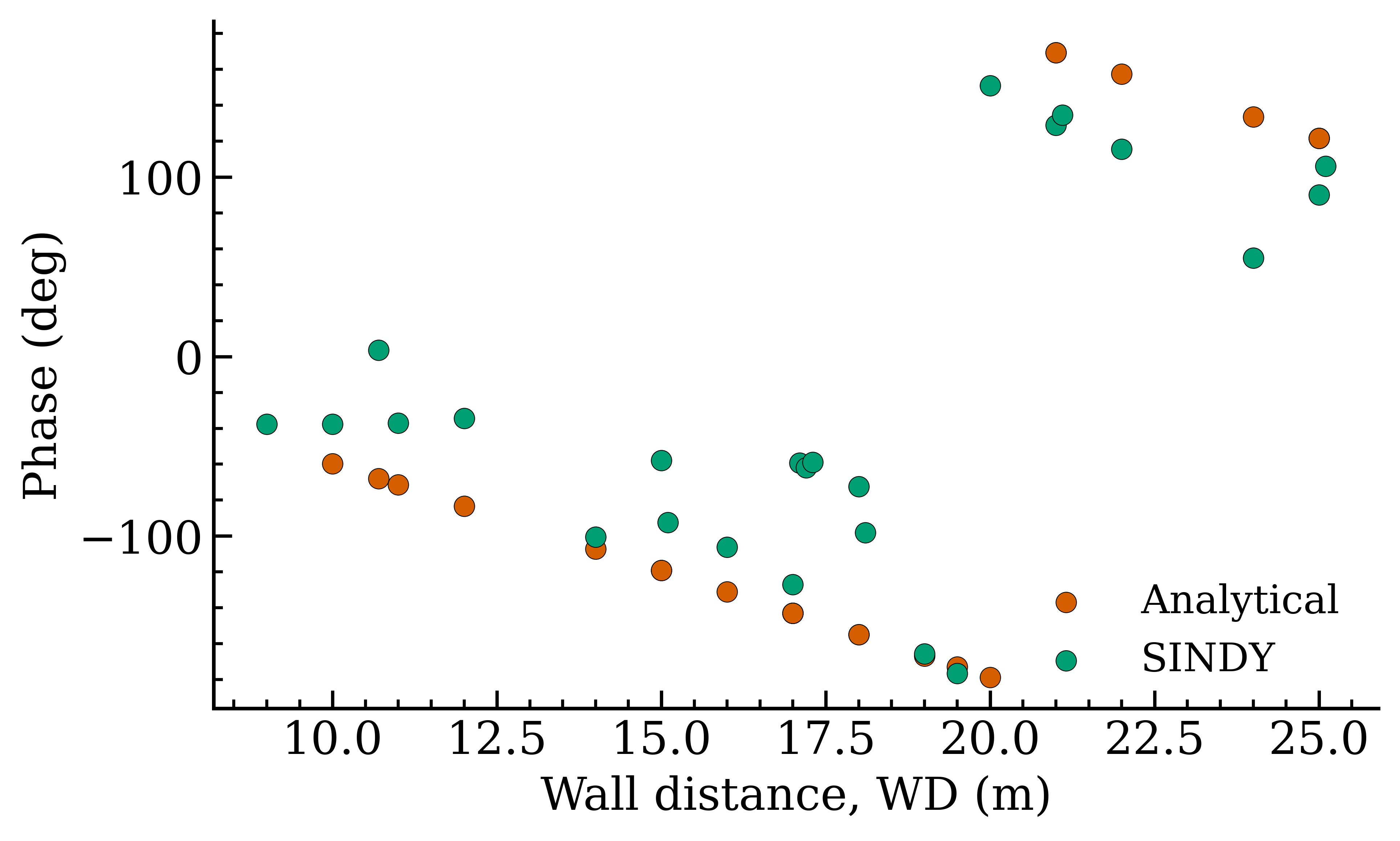} 
    \caption{SINDy identified excitation force phase vs analytical predictions}
    \label{fig:sindy-phase}
\end{figure}

The phase of the excitation force can be predicted from a simple first-principles derived wave travel time argument based on the WD and the wave celerity $(c)$:
\begin{equation}
    t_{\mathrm{delay}} = \frac{2WD}{c}, \quad \phi = \omega t_{\mathrm{delay}},    
    \label{eq:analyical-phase}
\end{equation}

and matches well with the phase extracted via SINDy in Figure \ref{fig:sindy-phase}.

\begin{figure}[t]
    \centering
    \includegraphics[width=\linewidth]{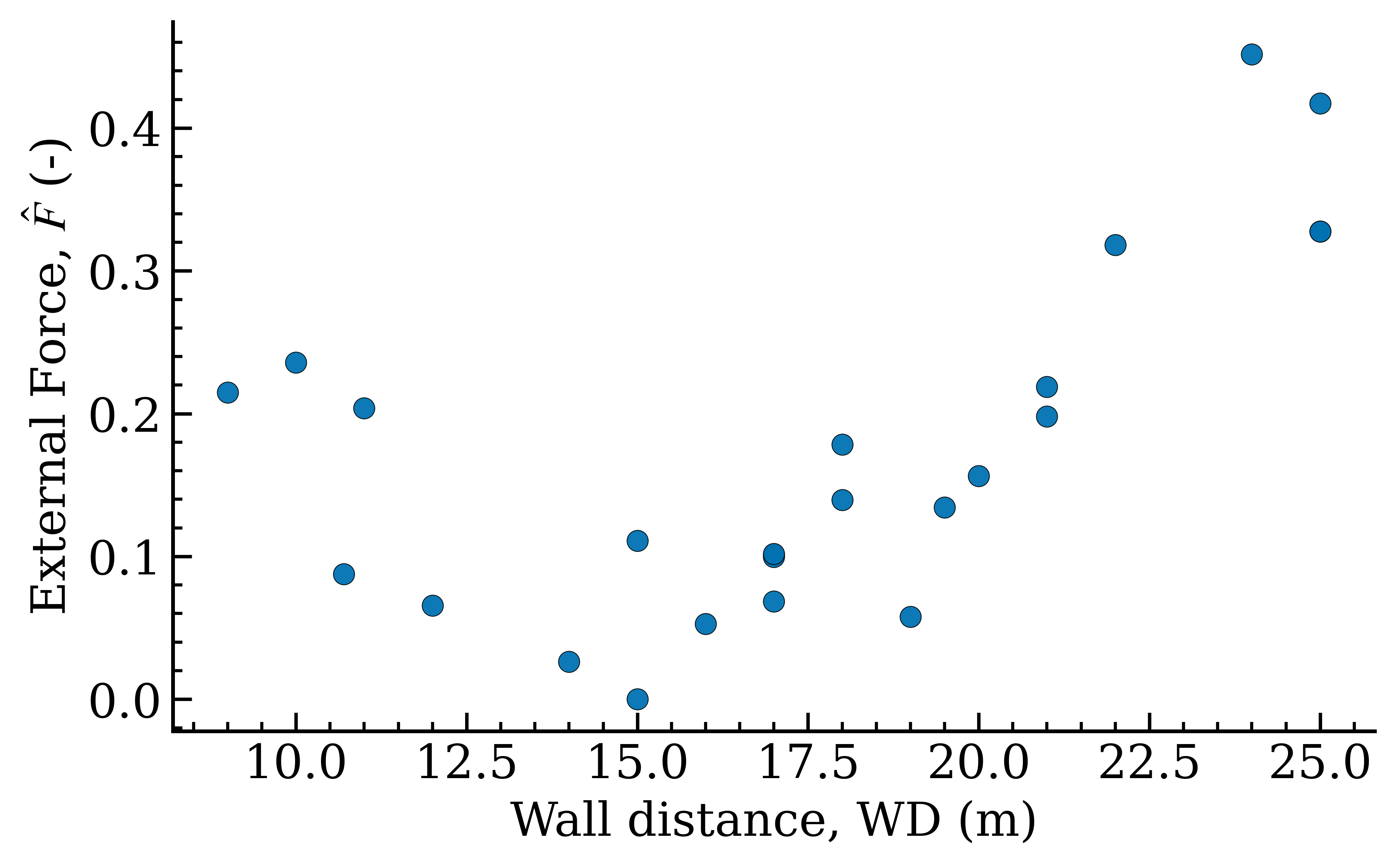} 
    \caption{SINDy identified excitation force magnitude}
    \label{fig:sindy-mag}
\end{figure}

SINDy identifies sinusoidal excitation terms whose phase aligns closely with the wave-travel-time prediction. Around WD=15 m, a minimal excitation forces amplitude is observed  (Figure \ref{fig:sindy-mag}), which is the same distance where the wave is 90 degrees offset from the spheres motion. No simple analytical relation for the excitation force magnitude has been found, but would likely require a similar methodology as \cite{Li2021}.

\subsection{SINDy-Informed Neural Network}

Sparse regression provides an interpretable ODE for each individual $(WD,DH)$ case, but the corresponding coefficient vectors ${\Xi}$ are only available at discrete points in the two-dimensional parameter space. The aim is to achieve a global surrogate valid across the full $(WD,DH)$ domain by determining a function that maps the coefficients over the full input domain. 

$${\Xi}(WD,DH)$$

This constitutes a novel neural-network formulation in which the network outputs the coefficients of a physically interpretable non-linear ODE, rather than the state trajectory itself, and these coefficients are explicitly confined to a SINDy-derived admissible range via coefficient rescaling in Equation \ref{eq:affine_rescale}.

A key design choice is that the network is trained to match the trajectory $x(t)$ rather than the coefficients themselves (Equation \ref{eq:loss}), reflecting the non-uniqueness of non-linear ODE representations, as multiple different coefficient matrices (${\Xi}$) can generate almost identical dynamics.

\begin{figure}[htbp]
    \centering
    \includegraphics[width=\linewidth]{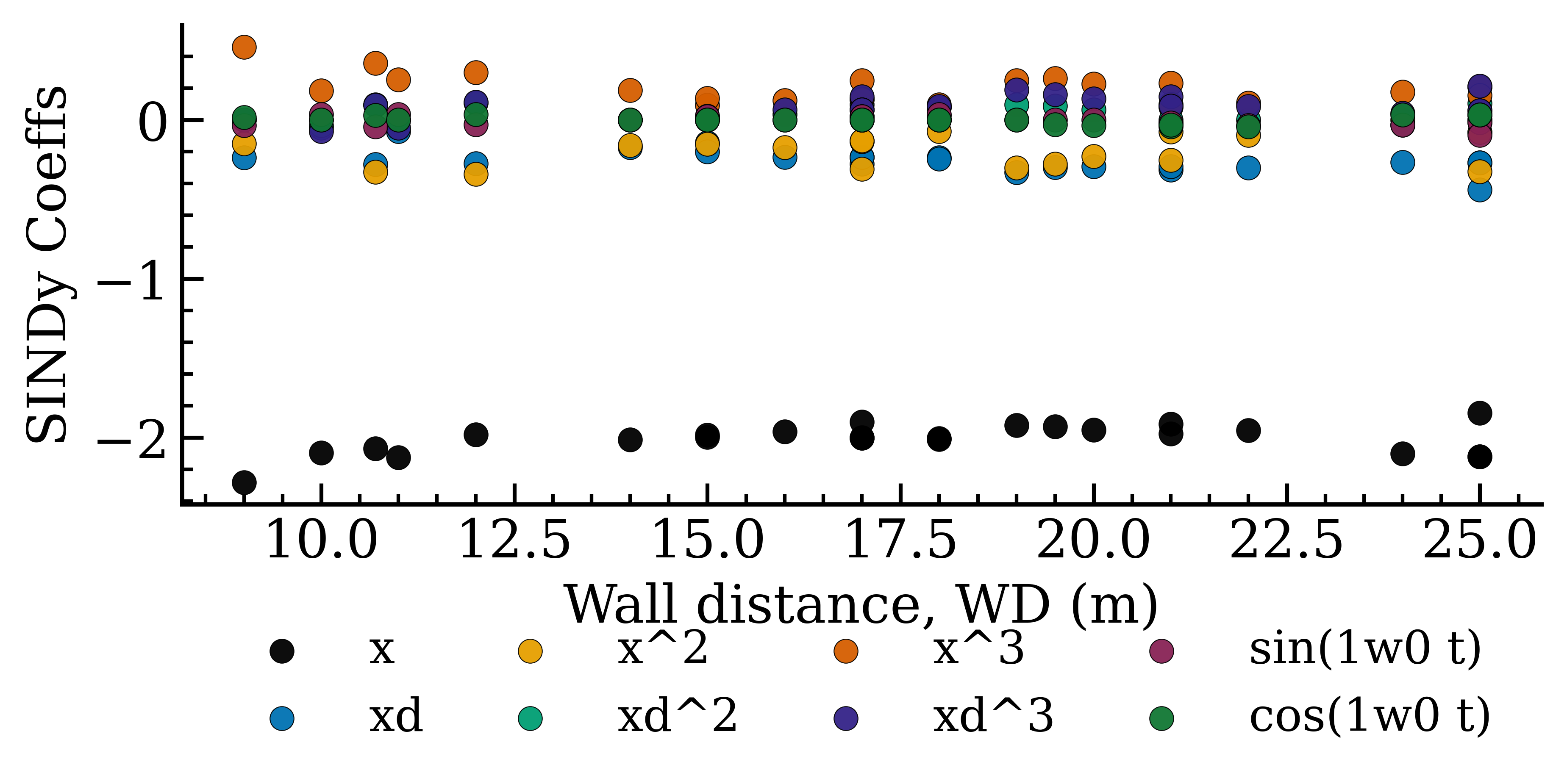}
    \caption{SINDy-identified coefficients.}
    \label{fig:sindy-coeffs}
\end{figure}

Figure \ref{fig:sindy-coeffs} shows a 2D slice for the magnitude of the SINDy-derived coefficients across all wall distances. This information is used as a prior within the NN to restrict the results to physically plausible ranges for each component of ${\Xi}$. 
\FloatBarrier

The neural network is implemented as a fully connected feed-forward network such that;
\begin{equation}
\mathcal{N}_\theta : \mathbb{R}^2 \xrightarrow{NN(WD,DH)} \mathbb{R}^8
\end{equation}

The network architecture used;
\begin{equation}
2(WD,DH) \xrightarrow{} 32 (W\times x^{l-1} +b_i)
\xrightarrow{tanh(\cdot)} 8(\Xi_{normalised} )
\end{equation}
consists of a single hidden layers of width 32 and a $\tanh$ non-linear activation. The final $\tanh$ layer yields output values from the neural networks such such that;
\[
{{\Xi}_{normalised}} \in [-1,1]^8,
\]
\begin{figure}[htbp]
    \centering
    \includegraphics[width=\linewidth]{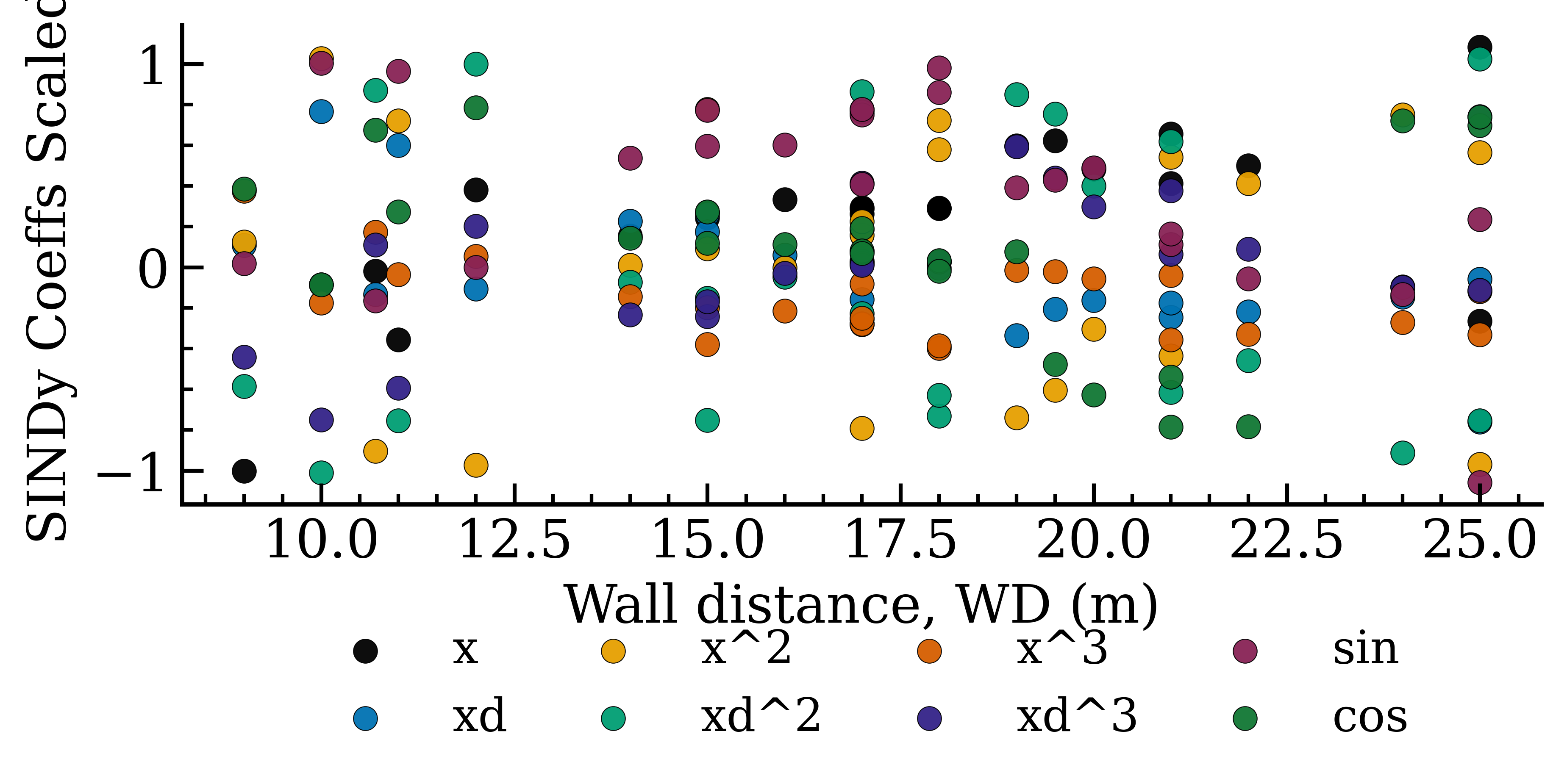}
    \caption{SINDy-identified coefficients normalised between $(-1,1)$.}
    \label{fig:sindy-coeffs-norm}
\end{figure}

After the output from the NN, in Figure \ref{fig:sindy-coeffs-norm}, the values are rescaled using SINDy-derived bounds:
\begin{equation}
\Xi_i
=
\frac{\xi_{\max}^i - \xi_{\min}^i}{2}\,{\Xi}_i
+ 
\frac{\xi_{\max}^i + \xi_{\min}^i}{2},
\qquad i = 1,\dots,8.
\label{eq:affine_rescale}
\end{equation}
Here, $(\xi_{\min}^i,\xi_{\max}^i)$ are chosen so that the resulting $\Xi_k$ span the range of SINDy-identified coefficients over all CFD trajectories. Equation~\eqref{eq:affine_rescale} therefore acts as a prior, constraining the learned coefficients to plausible hydrodynamic values.

Given a set of $N_\mathrm{tr}$ CFD trajectories, the NNs parameters $W_k, b_k$ are obtained by minimising the mean-squared error between the predicted and CFD responses,
\begin{equation}
MSE_{loss}
=
\frac{1}{N_\mathrm{tr}}
\sum_{i=1}^{N_\mathrm{tr}}
\frac{1}{T_\mathrm{s}}
\sum_{j=0}^{T_\mathrm{s}-1}
\left[
x_\theta^{(i)}(t_j) - x^{(i)}(t_j)
\right]^2
\label{eq:loss}
\end{equation}
where $x_\theta^{(i)}(t_j)$ is obtained by numerically integrating using RK4 (Equation \ref{eq:RK4}) the non-linear ODE with coefficients ${\Xi} = \mathcal{N}_\theta(WD^{(i)},DH^{(i)})$. The loss \eqref{eq:loss} is minimised using the Adam optimiser.

The test loss is monitored to detect overfitting. Once trained, the surrogate model can be evaluated at arbitrary $WD, DH$ values, providing an efficient surrogate for the CFD solver.

In summary, this framework involves sparse regression based ROMs followed by neural network based interpolation of the change in hydrodynamic coefficients with initial conditions. SINDy enforces interpretability, and provides the parametric structure, while the neural network supplies a smooth, data-driven coefficient manifold over parameter space (Figure \ref{fig:Manifold}), yielding a real-time physics-constrained surrogate model for the heave decay near a wall.

\subsection{Training}
The network training utilises a standard PyTorch optimisation loop. However, the computational efficiency is constrained by the nature of the loss function, which requires numerical integration of the identified system of equations. The numerical integration is implemented sequentially, preventing effective parallelisation on a GPU. Consequently, the forward and backward passes are executed primarily in series on the CPU, resulting in a training duration of approximately 24 hours. 

To ensure numerical stability when re-integrating the NNs identified ODE, a constrained initialisation strategy was implemented. The weights and biases were stochastically initialised within a narrow range to keep predictions physically plausible. These constraints were then gradually relaxed, effectively preventing divergence during the first few training epochs.

\begin{figure}[ht]
    \centering
    \includegraphics[width=0.9\linewidth]{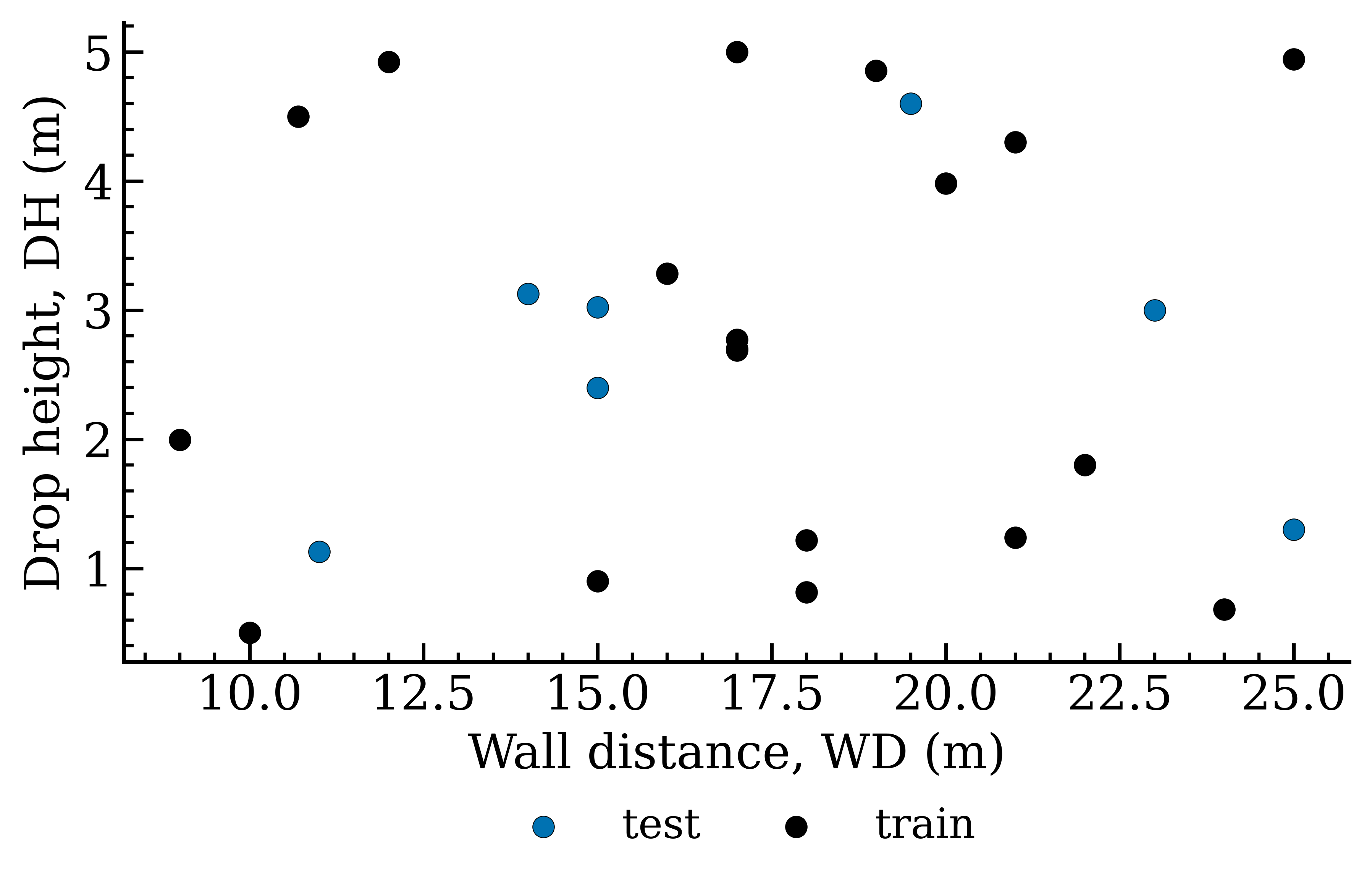}
    \caption{Training data (blue) and testing data (black).}
    \label{fig :Extrap /interp}
\end{figure}

\section{Results}
\subsection{Surrogate Accuracy}
\begin{figure}
    \centering
    \includegraphics[width=0.8\linewidth]{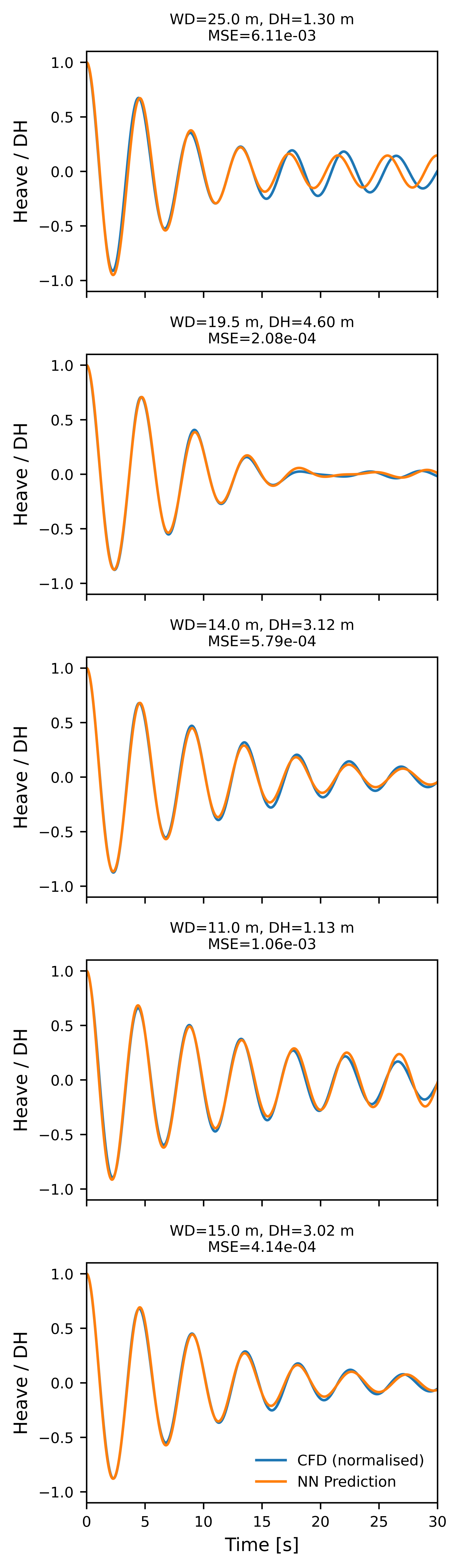}
    \caption{Neural-network prediction vs CFD for test cases.}
    \label{fig:NN-results}
\end{figure}

The surrogate model demonstrates high predictive fidelity (Figure \ref{fig:NN-results}). The maximum achievable accuracy over the domain, based on the idealised SINDy-derived coefficients, is approximately $5.1 \times 10^{-4}$ (see Figure~\ref{fig:Coefficient-library}). In comparison, the neural network achieves an average training and testing (interpolation-set) error of
\[
\mathrm{MSE}_{\mathrm{train}} \approx 4.85\times 10^{-4}, \qquad
\mathrm{MSE}_{\mathrm{test}} \approx 5.76\times 10^{-4}.
\]
Here, the reported testing error excludes the extrapolated case at $WD=25.0\,\mathrm{m},\ DH=1.30\,\mathrm{m}$, which falls just outside outside the black markers that make up the training domain in Figure \ref{fig :Extrap /interp}.

Interestingly, training error falls slightly below the average SINDy identified model error of $\approx 5.0e-4$ from Figure \ref{fig:Coefficient-library}. In conjuction with the fact that the NN identified manifolds do not intersect the SINDy identified coefficients, this indicates that the neural network is not replicating the linear-algebra-derived coefficients, but is instead learning a different representation that yields trajectories more consistent with the raw data. 

\begin{figure*}[t!]
    \centering
    \includegraphics[width=6in]{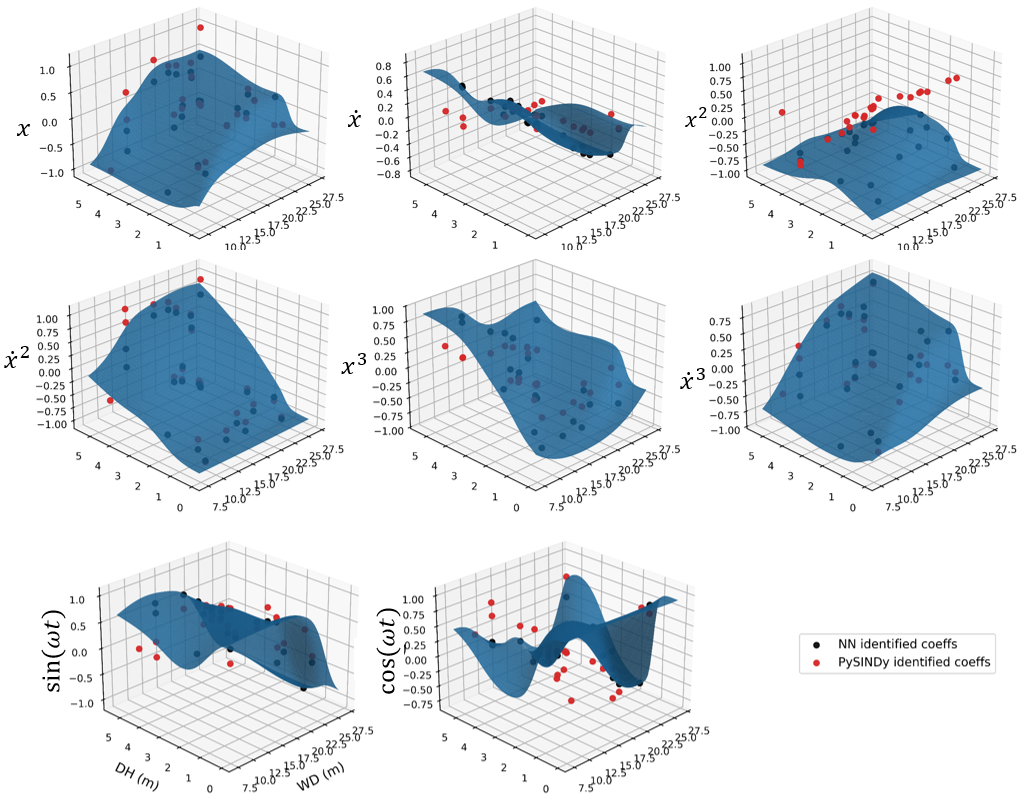}
    \caption{Learned coefficient manifolds}
    \label{fig:Manifold}
\end{figure*}

\section{Discussion}
\subsection{Interpolation and Extrapolation}

Figure~\ref{fig :Extrap /interp} shows which validation and test points lie within the interpolation region and which correspond to extrapolation. As expected, interpolation yields more accurate trajectories, while extrapolated conditions produce noticeably larger errors.





\subsection{Neural Network Outputs}
The neural network–identified coefficient manifolds are shown in Figure ~\ref{fig:Manifold}. These surfaces represent the NN's prediction throughout the entire input space with the training dataset, represented by the blue markers, lying on this manifold. This confirms that the NN has successfully learned a smooth parametric representation of the coefficient.

The neural network's smoothed coefficient manifolds deviates significantly from those obtained by direct linear algebra through SINDy in Figure \ref{fig:Manifold}. However, both sets generate nearly identical dynamical responses. The mismatch is likely due to the lack of a $L_2$ regularisation parameter $\alpha$ within the NN optimisation as apposed to ridge regression. Furthermore, we hypothesise that this is due to the fact that for non-linear systems, there is no singular best solution. Note that the neural network identifies a set of coefficients that minimises trajectory error rather than the error to the SINDy derived coefficient.

\section{Conclusion}

The results demonstrate that interpretable non-linear ODEs can be reconstructed from CFD data using sparse regression, and that these parametric coefficients can be embedded into a NN to form an efficient reduced-order surrogate model. Across the sampled input domain, the SINDy-based identification consistently recovers hydrostatic and radiation contributions that agree with analytical expectations and with previously validated linear and weakly non-linear models for the heaving sphere. In particular, the close agreement between analytically derived and regression-based coefficients for the linear and cubic stiffness terms confirms that the dominant non-linear hydrostatic mechanisms are well captured. This gives confidence that the learned surrogate is anchored to physically meaningful force components rather than arbitrary empirical fits.

By contrast, the excitation terms associated with reflected waves are found to be substantially more complex. The failure of a linear decomposition strategy to isolate excitation force may indicate that the excitation force may not be treated as state-independent forcing. The simple single-frequency $\sin/\cos$ terms included in the SINDy library provide only a crude surrogate for these effects. They recover the correct phase delay, which is consistent with a wave travel-time argument based on $2WD/c$, but they do not reproduce the true excitation force realistically. This underscores the need for more sophisticated forcing representations if diffraction-dominated regimes are to be modelled with higher fidelity, and allow the learned coefficients to reproduce a higher accuracy representation of the data.

The SINDy-informed neural network (NN) offers an interpretable method of designing the surrogate model. Rather than attempting to learn the dynamics directly from trajectories in a black-box fashion, the NN acts on a low-dimensional space of coefficients. 

The analysis on interpolation and extrapolation further illustrate the strengths and limitations of the approach. Within the training data, the network attains mean-squared errors close to the theoretical minimum found though applying SINDy on the testing data, indicating that the mapping $(WD,DH)\mapsto{\Xi}$ has been learned with high accuracy.

From an application perspective, the proposed ROM provides a useful building block for launch-and-recovery analysis. It replaces expensive CFD runs with a real-time surrogate that can evaluate new $WD,DH$ configurations, while retaining a clear link between the model coefficients and underlying hydrodynamic mechanisms (added mass, damping, stiffness, and excitation). 

Future work may also incorporate uncertainty estimates in the parametrised ODE model. Furthermore, it is clear to see how this methodology can be extended to parameterise other key characteristics such as the sphere density, centre of mass and incident wave characteristics, but as the input grows in size, so too will the required training data grow proportionally. 

At the same time, several open questions remain. The present work considers only a highly idealised geometry and a single mode of motion (heave), and it focuses on regularised decay responses rather than fully irregular seas or six-degree-of-freedom dynamics. Extending the methodology to more complex hull forms, coupling modes of motion will require both richer libraries of candidate non-linear terms and more robust NN architectures. It is foreseen that the addition of extra physics, for the higher degree of freedom model, with a more complicated geoemtry will require a more expansive and thought out candidate library of terms. Additionally, while the current formulation uses CFD as the sole source of training data, the framework is compatible with experimental measurements, which could be incorporated in future.

\FloatBarrier

\printbibliography

\end{document}

\subsection{NN Interpretability}

- 3rd order damping
-3rd order hydrostatic
- NN identified Sin/cosine
By plotting the NN identified coefficients that relate the hydrostatic force, we can see that t
\begin{figure}[ht]
\centering
\includegraphics[width=\linewidth]{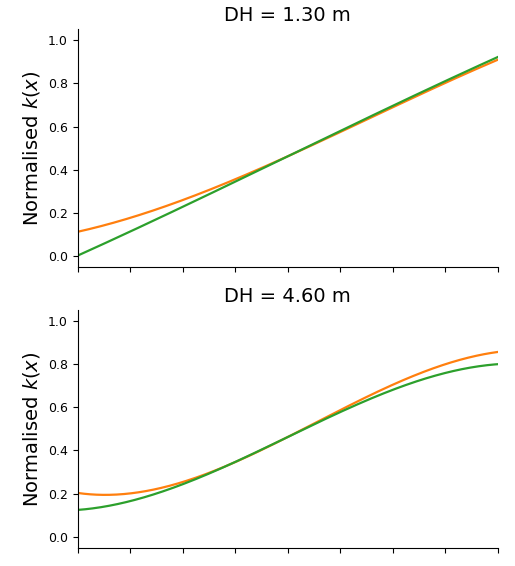}
\caption{NN identified restoring force.}
\label{fig:K-interp}
\end{figure}
\FloatBarrier

\begin{figure}[ht]
\centering
\includegraphics[width=\linewidth]{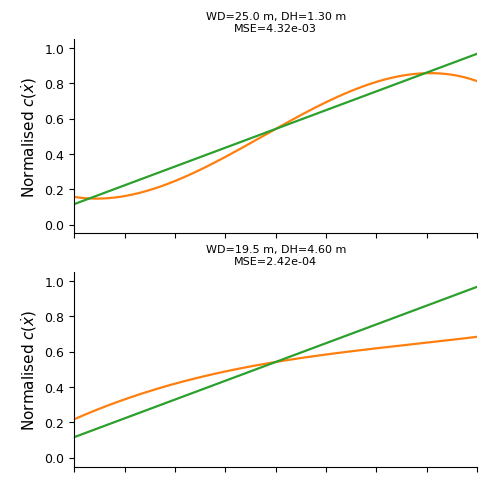}
\caption{NN identified damping force.}
\label{fig:K-interp}
\end{figure}
\FloatBarrier

\begin{figure}[ht]
\centering
\includegraphics[width=\linewidth]{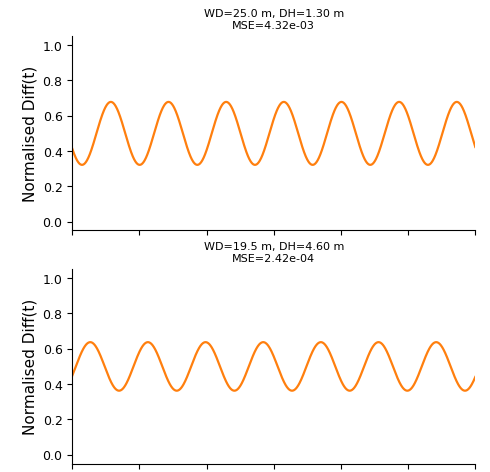}
\caption{NN identified excitation force.}
\label{fig:K-interp}
\end{figure}
\FloatBarrier